\documentclass[reprint,aps,jcp,amsmath,amssymb,english]{revtex4-2}
\usepackage{babel}
\usepackage{cancel}
\usepackage{chemformula}
\usepackage{bm}
\usepackage{float}
\usepackage{multirow}
\usepackage{booktabs}
\usepackage{rotating}
\usepackage{siunitx}
\usepackage{xcolor}
\usepackage{braket}
\usepackage{verbatim}
\usepackage[normalem]{ulem}
\begin{document}
\title{Second-Order Magnetic Properties in Paramagnetic Molecules From a Current Density Formulation
Including Scalar Relativistic Effects}
	\author{Francesco Ferdinando Summa}
	\email{fsumma@unisa.it}
	\affiliation{Dipartimento di Chimica e Biologia ``A. Zambelli'', Universit\`{a} degli Studi di Salerno, via Giovanni Paolo II 132, Fisciano 84084, SA, Italy}
	\author{Sonia Coriani}
	\affiliation{DTU Chemistry, Technical University of Denmark, Kemitorvet Bldg. 207, DK-2800 Kongens Lyngby, Denmark}
    \author{Andr\'e Severo Pereira Gomes}
	\affiliation{CNRS, UMR 8523–PhLAM–Physique des Lasers Atomes et Molécules, Univ. Lille, F-59000 Lille, France}
\begin{abstract}
	This work presents the theoretical background for the computation of nuclear magnetic shielding, nuclear hyperfine  and magnetizability tensors of paramagnetic molecules, using a magnetically induced current density framework to account for both orbital and spin contributions.  
    The resulting magnetizability tensor is fully consistent with the general Van Vleck formulation, recovering the temperature-dependent Curie contribution through the explicit integration of the magnetically induced spin current density. The methodology proposed herein provides a straightforward computational route that bypasses the complex evaluation of g-tensors and Zero-Field Splitting (ZFS) Hamiltonians. While the theoretical framework is general, we present applications rooted on physically motivated approximations where scalar relativistic effects are incorporated through corrections based on the Zeroth-Order Regular Approximation (ZORA) Hamiltonian within the ground-state spin density. This approach combines a relativistic self-consistent field (SCF) calculation for the ground-state spin density with a non-relativistic,  origin-independent current density calculation for the orbital contribution.  This hybrid strategy is shown to capture the  Heavy-Atom Light-Atom (HALA) effect in $^{1}\text{H}$ and $^{13}\text{C}$  shieldings, particularly in paramagnetic
    molecular systems containing transition metals up to the 3d series. By restricting the relativistic treatment  to the spin density, where scalar relativistic effects are dominant, and  neglecting such effects on the orbital contribution of light atoms, this method  offers a good compromise between computational efficiency and accuracy 
    for the characterization of large open-shell molecular systems.
    
\end{abstract}
	\maketitle
    
\section{Introduction}

The characterization of open-shell systems via Nuclear Magnetic Resonance (NMR) spectroscopy remains one of the most challenging tasks in computational chemistry, primarily due to the complex interplay between orbital and spin-dependent interactions. In paramagnetic molecules, the presence of unpaired electrons introduces large, temperature-dependent paramagnetic NMR (pNMR) shifts that are highly sensitive to both the local electronic environment and relativistic effects~\cite{Novotny2024}. 

Traditional approaches for calculating second-order magnetic properties (shieldings, magnetizabilies and nuclear hyperfine tensors) often rely on the evaluation of $g$-tensors and Zero-Field Splitting (ZFS) Hamiltonians, which can be computationally demanding and sensitive to the computational setup (e.g.\ require high-quality integration grids), especially for larger molecular systems.  
An alternative and powerful route is offered by the magnetically induced current density framework. This approach provides a rigorous spatial mapping of the electronic response to external magnetic perturbations, allowing for a more intuitive understanding of the magnetic properties. 
Current density methods are well-established for diamagnetic species, and their extension 
to open-shell systems was pioneered by~\citet{soncini_charge_2007}. 

In this work we generalize Soncini's framework to relativistic Hamiltonians, and demonstrate that this generalization is physically equivalent to the formalisms of~\citet{pennanen_density_2005} and~\citet{franzke_paramagnetic_2024}, provided that scalar relativistic effects are properly incorporated within the ground-state spin density. 
Furthermore, we show that the magnetizability tensor expression derived from our current density approach is fully consistent with the general Van Vleck formulation. By explicitly integrating the magnetically induced spin current density, we successfully recover the temperature-dependent Curie contribution without requiring the complex evaluation of intermediate $g$ and ZFS tensors.  

Although the theoretical framework introduced here is general, we illustrate its applicability through
a first implementation in the SYSMOIC~\cite{monaco_program_2021} code based on a 
Zeroth-Order Regular Approximation (ZORA) formulation~\cite{van_lenthe_relativistic_1994} of the current density. 
Unlike perturbative methods such as Breit-Pauli~\cite{mcweeny_methods_1992}, which may suffer from singularities at the nuclear positions, 
the ZORA approach regularizes the kinetic energy operator, ensuring a robust description of the electronic structure near the nuclei. This is particularly relevant for capturing the Heavy-Atom Light-Atom (HALA) effect in $^1$H and $^{13}$C shieldings, where relativistic contributions from the heavy center are essential for an accurate description of the light-atom chemical shifts. 

We note that this first implementation deliberately focuses on spin-dependent contributions arising from the interaction with the external magnetic fields, as these exhibit the highest sensitivity to relativistic effects in paramagnetic systems. In contrast, the equivalent orbital contributions are 
currently treated within a standard non-relativistic scheme.
While this approximation (referred to as \textit{field-free spin-orbit coupling (SOC) response approximation})
results in large errors for heavier atoms, where SOC is significant and can profoundly alter the topology 
and magnitude of magnetically induced currents, as recently demonstrated for  heavy-atom hydrides via trans-ligand pathways \cite{Blasco2026}, 
we show 
that, for systems up to 3d transition metals, scalar relativistic corrections are often sufficient for a wide range of molecular systems.
Consequently, adopting a scalar 
relativistic treatment exclusively for the spin density provides the pNMR community with a robust and computationally efficient route for the characterization  of large-scale open-shell systems.
Moreover, by leveraging magnetically induced current densities, we bridge to the methodologies and protocols established for closed-shell molecules using non-relativistic Hamiltonians. 

The manuscript is organized as follows: 
Section~\ref{Theoretical_Approach} provides a theoretical overview of the model based on induced current densities, with the specific formulations for the nuclear magnetic shielding, nuclear hyperfine coupling, and magnetizability tensors 
detailed in Section~\ref{Current_Density_Formulations}. 
Section \ref{Implementation} and \ref{Results_and_Discussion} illustrate the implementation of the proposed model at the GHF (Generalizes Hartree-Fock), GKS (Generalized Kohn-Sham), HF (Hartree-Fock), and DFT 
(Density Functional Theory) levels of theory, alongside several reference calculations to demonstrate its consistency.
Section~\ref{Conclusions} summarises our conclusions.

\section{Theoretical Approach}
\label{Theoretical_Approach}

For a molecule with $n$ electrons and $N$ clamped nuclei, charge, mass,
	position, canonical and angular momentum of the $k$-th 
	electron are indicated, in the configuration space, by 
	$-e$, $m_{\textrm{e}}$, ${\boldsymbol{r}}_k$,
	$\hat{\boldsymbol{p}}_k$, 
	$\hat{\boldsymbol{l}}_k={\boldsymbol{r}}_k\times\hat{\boldsymbol{p}}_k$,
	$k=1,2\ldots n$, using boldface letters for electronic operators. Analogous quantities for nucleus $I$ 
	are $Z_Ie$, $M_I$, ${\boldsymbol{R}}_I$, etc.  for $I=1,2\ldots N$.
	The imaginary unit is represented by a Roman $ {\textrm{i}} $.
	Throughout this paper, SI units are used and standard tensor formalism is employed, e.g.,
	the Einstein convention of implicit summation over two repeated Greek indices is in force.
	The third-rank pseudotensor defined by Ricci and Levi-Civita is indicated by $\epsilon_{\alpha\beta\gamma}$.
	Capitals denote $ n $-electron vector operators, e.g., the operator 
	representing the electric field  acted by the $k$-th electron upon the $I$-th nucleus is expressed by
	\begin{equation}
		\label{EkIfield}
		\hat{\boldsymbol{E}}_{I}^{k}   =\frac{e}{4\pi\epsilon_{0}} \frac{\boldsymbol{r}_{k}-\boldsymbol{R_{I}}}
		{|\boldsymbol{r}_{k}-\boldsymbol{R_{I}}|^{3}},
		\quad 
		\hat{\boldsymbol{E}}_{I}^{n}=\sum_{k=1}^n \hat{\boldsymbol{E}}_{I}^{k},
	\end{equation}
	and the corresponding operator
	for the electric field 
    exerted by nucleus $I$ on the
	$k$-electron  
    is
\begin{equation}
\hat{\boldsymbol{E}}_{k}^{I}=Z_{I}\hat{\boldsymbol{E}}_{I}^{k}~.
\end{equation}
To handle spin effects in the correct phenomenological way, our starting point is the Dirac Hamiltonian for 
one electron in a non-vanishing electromagnetic field
\begin{equation}
\hat{H}_{D}=c\boldsymbol{\alpha}\cdot\hat{\boldsymbol{\pi}}+\boldsymbol{\beta}m_{\textrm{e}}c^2+\left[\hat{V}-e\Phi\right]\boldsymbol{I}_4
	\label{DiracHam}
\end{equation}
where
\begin{equation}
\hat{\boldsymbol{\pi}}=\hat{\boldsymbol{p}}+e\boldsymbol{A}
\end{equation}
is the mechanical momentum operator of the particle, $\boldsymbol{A}$ and $\Phi$ denote the magnetic vector and scalar potentials, respectively, and $\hat{V}$ represents the electrostatic potential. Moreover,
\begin{equation}
	\boldsymbol{\alpha}=\begin{pmatrix}
		0 & \hat{\boldsymbol{\sigma}} \\
	\hat{\boldsymbol{\sigma}}& 0
	\end{pmatrix}
	\qquad
	\boldsymbol{\beta}=\begin{pmatrix}
		\boldsymbol{I}_2 & 0 \\
		0 & -\boldsymbol{I}_2  
	\end{pmatrix}
\end{equation}
where $\boldsymbol{I}_n$ is an identity matrix of dimensions $n\times n$, and
$\boldsymbol{\hat{\sigma}}$ is a vector operator collecting the Pauli matrices
\begin{equation}
	\label{Paulimats}
	\hat{\sigma}_x=\left(\begin{array}{ll}
		\,0 & \,1 \\
		\,1 & \,0
	\end{array}\right)~; \quad 	\hat{\sigma}_y=\left(\begin{array}{rr}
		\,0 & -\mathrm{i} \\
		\,\mathrm{i} & \, 0
	\end{array}\right)~; \quad 	\hat{\sigma}_z=\left(\begin{array}{rr}
		\,1 & \,0 \\
		\,0 & -1
	\end{array}\right)~.
\end{equation} 
The eigenvectors of $\hat{\sigma}_z$ are commonly indicated by
\begin{equation}
	\nonumber
	\label{alfabeta}
	\vert \alpha \rangle \equiv
	\begin{pmatrix} 
		1
		\\
		\thinspace 0
	\end{pmatrix}~;
	\quad
	\vert \beta \rangle \equiv
	\begin{pmatrix} 
		0
		\\
		\thinspace 1
	\end{pmatrix}~.
\end{equation}
To obtain a simpler description of the Hamiltonian in Eq.~\eqref{DiracHam}, one can transform its 4-component representation into a block diagonal form by applying the Foldy-Wouthuysen transformation~\cite{bjorken1964relativistic}.
This approach reduces the problem to an effective 2-component form. 
For problems of chemical interest, one can then focus on the large-component wavefunctions, which dominate the positive-energy solutions.
For the hydrogen atom (i.e., a Coulomb-like potential), this technique yields~\cite{mcweeny_spins_1970,mcweeny_methods_1992}
	\begin{equation}
			\begin{split}
		\hat{H}_{D}\simeq &\,\,m_{\textrm{e}}c^2\boldsymbol{I}_2+ \frac{\hat{\boldsymbol{\pi}}^2}{2m_{\textrm{e}}}\boldsymbol{I}_2+g_{\textrm{e}}\frac{\mu_{B}}{\hbar}\hat{\boldsymbol{s}}\cdot\left[\boldsymbol{\nabla}\times\boldsymbol{A}\right]\\
		&+\left[\hat{V}-e\Phi\right]\boldsymbol{I}_2+\frac{e}{2m_{\rm e}^2c^2}\hat{\boldsymbol{s}}\cdot\left[\hat{\boldsymbol{E}}\times\hat{\boldsymbol{\pi}}\right]+\cdots
	\end{split}
		\label{breit-pauli-2}
	\end{equation}
	that is, an approximated Breit-Pauli Hamiltonian where 
	$\mu_{B}$ is the Bohr magneton, $g_{\textrm{e}}$ is the electron spin g-factor and
	\begin{equation}
\boldsymbol{\hat{s}}=\frac{\hbar}{2}\hat{\boldsymbol{\sigma}}
    \label{eq:spin-operator}
	\end{equation}
	is the spin operator. Note that scalar relativistic corrections, such as the Darwin and mass-velocity terms, have been omitted from Eq.~\eqref{breit-pauli-2} for sake of clarity. In the presence of an external magnetic field, the mass-velocity term would involve the fourth power of the mechanical momentum operator, $\hat{\boldsymbol{\pi}}^4$, leading to complex higher-order couplings between the magnetic field and the electron momentum.

    The generalization of the Breit-Pauli Hamiltonian in Eq.~\eqref{breit-pauli-2} to the case of a molecular system is straightforward. Indeed, within the Born--Oppenheimer approximation,  for applied static and uniform magnetic and electric fields, we have
\begin{equation}
	\begin{split}
		\hat{H}= &\frac{1}{2m_{\textrm{e}}}\sum_{k}^{n}\hat{\boldsymbol{\pi}}_k^{2}\boldsymbol{I}_2+\underbrace{g_{\textrm{e}}\frac{\mu_{B}}{\hbar}\sum_{k}^{n}\hat{\boldsymbol{s}}_k \cdot \left[\boldsymbol{\nabla}\times\boldsymbol{A}_k\right]}_{\textrm{spin Zeeman}}
		\\&+\underbrace{\frac{e}{2m_{\textrm{e}}^2c^2}\sum_{k,I}^{n,N}\hat{\boldsymbol{s}}_k\cdot \left [\hat{\boldsymbol{E}}_{k}^{I}\times \hat{\boldsymbol{\pi}}_k\right ]}_{\textrm{spin-orbit coupling}}-e\sum_{k}^{n}\Phi_k\boldsymbol{I}_2\\&+\frac{e^2}{8\pi \varepsilon_{0}}\left[\sum_{k,j\neq k}^{n,n}  
        \frac{1}{r_{kj}}-2\sum_{I,k}^{N,n} \frac{Z_{I}}{r_{I k}}+\sum_{I,K\neq I}^{N,N}\frac{Z_{I} Z_{K}}{R_{IK}}\right]\boldsymbol{I}_2
	\end{split}
	\label{many hamiltonian}
\end{equation}
where $r_{kj}=|\boldsymbol{r}_k-\boldsymbol{r}_j|$ is the distance between the $k$-th and the $j$-th electron, $r_{Ik}=|\boldsymbol{R}_I-\boldsymbol{r}_k|$ is the distance between $I$-th nucleus and $k$-th electron and $R_{IK}=|\boldsymbol{R}_I-\boldsymbol{R}_K|$ is the distance between the $I$-th and the $K$-th nuclei.
Note that the energy scale has been shifted by subtracting the electron rest mass energy to facilitate the comparison between relativistic and non-relativistic results. 

In addition to the simplified expression for the one-electron spin-orbit coupling Hamiltonian (strictly valid for a Coulomb like potential) in Eq.~\eqref{many hamiltonian}, we can now introduce the spin density matrix $\boldsymbol{Q}(\boldsymbol{r} ;\boldsymbol{r}^{\prime})$, defined as
\begin{equation}
	\boldsymbol{Q}(\boldsymbol{r} ;{\boldsymbol{r}}^{\prime})=\int_{\eta^{\prime}_1=\eta_1}	\boldsymbol{\hat{s}}(1)\gamma\left(\boldsymbol{x}_{1} ; \boldsymbol{x}_{1}^{\prime}\right)d\eta_1
\end{equation}
where $\boldsymbol{x}=(\boldsymbol{r},\eta)$ is a combined spatial and spin electron coordinate
\begin{equation}
	\gamma\left(\boldsymbol{x}_1 ; \boldsymbol{x}_1^{\prime}\right)=n \int \Psi\left(\boldsymbol{x}_1, \boldsymbol{X}_1\right) \Psi^*\left(\boldsymbol{x}_1^{\prime}, \boldsymbol{X}_1\right) d \boldsymbol{X}_1
\end{equation}
and $d\boldsymbol{X}_1 \equiv\left\{d \boldsymbol{x}_2, \ldots, d\boldsymbol{x}_n\right\}$.
Equating $\boldsymbol{r}=\boldsymbol{r}^{\prime}$, we obtain the spin density, described by the axial vector
\begin{equation}
	\label{qgamma}
	\boldsymbol{Q}(\boldsymbol{r})\equiv \boldsymbol{Q}(\boldsymbol{r} ;\boldsymbol{r})~.
\end{equation}
By employing the Hamiltonian in Eq.~\eqref{many hamiltonian} and following the Landau approach based on the definition
\begin{equation}
	\delta {H}_{c}=-\int \,\boldsymbol{J}(\boldsymbol{r})\cdot\delta\boldsymbol{A}(\boldsymbol{r})\,d^3 r~,
	\label{classic H variation}
\end{equation}
where the classical Hamiltonian ${H}_c$ is identified with the expectation value of the quantum-mechanical Hamiltonian according to
\begin{equation}
	{H}_{c}=\langle H\rangle =\langle \Psi|\hat{H}|\Psi\rangle~, 
\end{equation}
it can be shown that,
within the Born-Oppenheimer and  strictly central field approximations, 
the total induced semi-relativistic electron current density for a generic open-shell system
is given by~\cite{summa_molecular_2024}
\begin{equation}
	\begin{split}
	\boldsymbol{J}(\boldsymbol{r})=&\underbrace{-\frac{e}{m_{\textrm{e}}}\Re\left[\hat{\boldsymbol{\pi}} \gamma\left(\boldsymbol{r};\boldsymbol{r}^{\prime}\right)\right]_{\boldsymbol{r}^{\prime}=\boldsymbol{r}}}_{\textrm{non-relativistic current}} \\&\underbrace{-g_{\textrm{e}}\frac{\mu_{B}}{\hbar}\,\boldsymbol{\nabla}\times\boldsymbol{Q}(\boldsymbol{r})
	-\frac{e^2}{2m_{\textrm{e}}^2c^2}\sum_{I=1}^{N}\boldsymbol{Q}(\boldsymbol{r})\times\hat{\boldsymbol{E}}^{I}}_{\textrm{spin-current}}
\end{split}
	\label{totalcurrent}
\end{equation}
where 
\begin{equation}
\hat{\boldsymbol{E}}^{I}=\sum_{k=1}^{n}\hat{\boldsymbol{E}}^{I}_{k}~.
\end{equation}
This expression differs from the one obtained using Gordon's decomposition~\cite{gordon_strom_1928} due to the presence of the spin-orbit coupling term, which appears as the last term in Eq. \eqref{totalcurrent}. 
This term is singular at the nuclear coordinates, see Eq. (\ref{EkIfield}), which poses challenges for its numerical evaluation in real space. 

The current density defined in Eq.~\eqref{totalcurrent} is by definition gauge-invariant for an exact calculation. 
In the SI system, units of $\boldsymbol{J}$ are $\left[\textrm{A}\,\textrm{m}^{-2}\right]$. 

The Hamiltonian in Eq.~\eqref{many hamiltonian} used to derive 
Eq.~\eqref{totalcurrent} describes the interaction of the electrons with the intramolecular perturbation. 
This perturbation arises from both the intrinsic intramolecular magnetic dipoles, $\boldsymbol{m}_I=\gamma_I\hbar\boldsymbol{I}_I=g_I\mu_{N}\boldsymbol{I}_I$, expressed via the magnetogyric ratio $\gamma_I$ and spin $\boldsymbol{I}_I$ of nucleus $I$  via the vector potential $\sum_{I=1}^{N} \mathbf{A}^{\boldsymbol{m}_I}$, and an external, spatially uniform and time-independent magnetic field $\boldsymbol{B}=\boldsymbol{\nabla}\times\boldsymbol{A}$,
\begin{align}
	&\boldsymbol{A}=\boldsymbol{A}^{\boldsymbol{B}}+\sum_{I}^{N} \boldsymbol{A}^{\boldsymbol{m}_I} \\
	&\boldsymbol{A}^{\boldsymbol{B}}=\frac{1}{2} \boldsymbol{B} \times \boldsymbol{r} \\
	&\boldsymbol{A}^{\boldsymbol{m}_I}=\frac{\mu_0}{4 \pi} \frac{\boldsymbol{m}_I \times\left(\boldsymbol{r}-\boldsymbol{R}_I\right)}{\left|\boldsymbol{r}-\boldsymbol{R}_I\right|^3}
\end{align}
In tensorial notation, the continuity equation associated with the total current density vector defined in Eq.~\eqref{totalcurrent} is
\begin{equation}
	\nabla_{\alpha}J_{\alpha}(\boldsymbol{r})=0
	\label{conteq}
\end{equation}
as illustrated in 
Ref.~\citenum{summa_molecular_2024}. This condition is fully satisfied only if the state functions are exact eigenfunctions of a model Hamiltonian and therefore satisfy the off-diagonal hypervirial theorem for the position operator, i.e. in HF, DFT, Full Configuration Interaction (FCI) or in other variational approaches \cite{epstein_variation_1974,summa_assessment_2021}. 
Furthermore, the condition 
is compatible with the true induced relativistic current density \cite{greiner_field_1996}. 
In practical application of 
Eq.~\eqref{totalcurrent}, this condition is always satisfied for the spin magnetization  and the spin-orbit coupling electron currents---the first and second contribution to the spin current term, respectively---but not for the non-relativistic one that exhibits gauge-dependence in calculations with finite basis sets~\cite{summa_molecular_2024}. 

The approximated Breit-Pauli Hamiltonian introduced in 
Eq.~\eqref{many hamiltonian} is not suitable for variational calculations, since the spin-orbit coupling term is variationally unstable due to the presence of a $r^{-3}$ singularity in its definition.

\subsection{Relativistic Formulation and Approximations}

As mentioned in the previous section, the expression for the current density in Eq.~\eqref{totalcurrent} is not a suitable starting point to properly incorporate scalar relativistic or spin-orbit coupling effects. To account for these contributions,\cite{sundholm_current_2021} a consistent relativistic framework 
must be employed,
such as the zeroth-order regular approximation (ZORA)~\cite{van_lenthe_relativistic_1994,van_lenthe_zero-order_1996,van_wullen_molecular_1998,van_lenthe_geometry_1999,bouten_relativistic_2000,dyall_introduction_2007,autschbach_chapter_2009},  Douglas-Kroll-Hess (DKH)~\cite{douglas_quantum_1974,hess_relativistic_1986,reiher_exact_2004}, or exact two-component (X2C)~\cite{kutzelnigg_quasirelativistic_2005,ilias_infinite-order_2007,x2c:2007}. 
We refer the interested reader to  Ref.~\cite{bouten_relativistic_2000,autschbach_nmr_2015} for specific details on the calculation of magnetic properties within ZORA, 
and to Refs.~\cite{kutzelnigg_relativistic_2009,franzke_reducing_2023} for X2C.

Due to its closed analytical form and its simplicity compared to DKH and X2C, we will use the ZORA approach to demonstrate our treatment. Specifically, by applying the Landau approach within the ZORA framework, we will obtain an analogous expression of the current density 
that simultaneously accounts for 
both 
scalar and spin-orbit relativistic effects.  

The ZORA Hamiltonian, in the presence of static and uniform magnetic and electric fields, can be written as:
\begin{equation}
	\hat{H}^{\textrm{ZORA}}=\sum_{k}^{n}\hat{T}^{\textrm{ZORA}}_k+\left[\hat{V}-e\Phi\right]\boldsymbol{I}_2
    \label{H_ZORA}
\end{equation}
where the ZORA kinetic energy is defined as
\begin{equation}
	\hat{T}^{\textrm{ZORA}}_k=\boldsymbol{\sigma}_k\cdot\hat{\boldsymbol{\mathbf{\pi}}}_k\,\mathcal{K}(\boldsymbol{r})\,\boldsymbol{\sigma}_k\cdot\hat{\boldsymbol{\mathbf{\pi}}}_k
	\label{ZORAEXP}
\end{equation}
and
\begin{equation}
	\mathcal{K}(\boldsymbol{r})=\frac{c^2}{2m_{\textrm{e}}c^2-eV(\boldsymbol{r})}
\end{equation}
is the ZORA relativistic scaling factor. Note that the ZORA Hamiltonian, as expressed in Eq.~\eqref{H_ZORA}, naturally incorporates the leading-order scalar relativistic effects, namely the mass-velocity correction and the Darwin term. 
The potential $V(\boldsymbol{r})$ in the denominator of $\mathcal{K}(\boldsymbol{r})$ is an effective electrostatic potential, see Appendix \ref{veff} for details regarding its evaluation. 

Applying the relation
\begin{equation}
\boldsymbol{\sigma}
\cdot
\hat{\boldsymbol{\mathbf{\pi}}}
\,
\mathcal{K}(\boldsymbol{r})\,\boldsymbol{\sigma}
\cdot
\hat{\boldsymbol{\mathbf{\pi}}}
=  
\left[\hat{\boldsymbol{\mathbf{\pi}}}
\,
\mathcal{K}(\boldsymbol{r})\right]
\cdot
\hat{\boldsymbol{\mathbf{\pi}}}
+ 
i\boldsymbol{\sigma}
\cdot
\left[\hat{\boldsymbol{\mathbf{\pi}}}
\,
\mathcal{K}(\boldsymbol{r}) \times \hat{\boldsymbol{\mathbf{\pi}}}\right]
\end{equation}
and substituting the scalar and vector potentials~\cite{dyall_introduction_2007}, 
we can rewrite $\hat{T}^{\textrm{ZORA}}_k$ (\ref{ZORAEXP}) as
\begin{equation}
	\begin{split}
\hat{T}^{\textrm{ZORA}}_k=&\left[\hat{\boldsymbol{\mathbf{\pi}}}_k\,\mathcal{K}(\boldsymbol{r})\,\hat{\boldsymbol{\mathbf{\pi}}}_k\right]\boldsymbol{I}_2+g_{\textrm{e}}\mu_{B}\frac{2m_{\textrm{e}}}{\hbar}\mathcal{K}(\boldsymbol{r})\hat{\boldsymbol{s}}_k \cdot \left[\boldsymbol{\nabla}\times\boldsymbol{A}_k\right]\\&+g_{\textrm{e}}\frac{e}{c^2}\mathcal{K}^2(\boldsymbol{r})\sum_{I=1}^{N}\hat{\boldsymbol{s}}_k\cdot \left [\boldsymbol{\nabla}V^{I}\times \hat{\boldsymbol{\pi}}_k\right]
\end{split}
\label{TZORA_1}
\end{equation}
in which the first term describes scalar contributions and the other two are 
spin contributions. By further expressing Eq.~\eqref{TZORA_1} in powers of the external magnetic field, we arrive at:
\begin{equation}
    \hat{T}^{\textrm{ZORA}}_k = \hat{h}_k^{(0)} + \hat{h}_k^{(1)} + \hat{h}_k^{(2)}
\end{equation}
where the zeroth-order term contains the scalar relativistic kinetic energy (first contribution in Eq.~\eqref{TZORA_1}) 
and the \textit{external magnetic field-free SOC} operator (third contribution in Eq.~\eqref{TZORA_1})
\begin{equation}
\hat{h}_k^{(0)} = \hat{\boldsymbol{p}}_k\mathcal{K}(\boldsymbol{r})\hat{\boldsymbol{p}}_k + g_{\textrm{e}}\frac{e}{c^2}\mathcal{K}^2(\boldsymbol{r})\sum_{I=1}^{N}\hat{\boldsymbol{s}}_k\cdot \left [\boldsymbol{\nabla}V^{I}\times \hat{\boldsymbol{p}}_k\right]~.
\end{equation}
The first-order perturbation, linear in the magnetic field, reads
\begin{equation}
\begin{split}
\hat{h}_k^{(1)} = \,\, & 2e\boldsymbol{A}_k\mathcal{K}(\boldsymbol{r})\hat{\boldsymbol{p}}_k + g_{\textrm{e}}\mu_{B}\frac{2m_{\textrm{e}}}{\hbar}\mathcal{K}(\boldsymbol{r})\hat{\boldsymbol{s}}_k \cdot \left[\boldsymbol{\nabla}\times\boldsymbol{A}_k\right] \\ 
& + g_{\textrm{e}}\frac{e^2}{c^2}\mathcal{K}^2(\boldsymbol{r})\sum_{I=1}^{N}\hat{\boldsymbol{s}}_k\cdot \left [\boldsymbol{\nabla}V^{I}\times \boldsymbol{A}_k\right]~,
\label{zora-linear-on-B}
\end{split}
\end{equation}
and contains both scalar and \textit{external magnetic field dependent spin contributions}. 
The second-order diamagnetic term is given by
\begin{equation}
\hat{h}_k^{(2)} = e^2\mathcal{K}
(\boldsymbol{r})\boldsymbol{A}_k^2~.
\label{zora-quadratic-on-B}
\end{equation}
For conciseness, in the following we will no longer explicitly indicate spin contributions as being independent and dependent to external magnetic fields, and rather use \textit{field-free} to distinguish the first from the second. Also, we note that the nuclear charge distribution model has to be employed in the ZORA Hamiltonian to avoid divergences in real space. 

From the ZORA Hamiltonian, one obtains the following expression for the current density 
\begin{widetext}
\begin{equation}
	\boldsymbol{J}(\boldsymbol{r}) = -2e \Re \left[ \mathcal{K}(\boldsymbol{r}) \hat{\boldsymbol{\pi}} \gamma(\boldsymbol{r};\boldsymbol{r}^\prime) \right]_{\boldsymbol{r}^\prime=\boldsymbol{r}} \underbrace{-g_{\textrm{e}}\mu_{B}\frac{2m_{\textrm{e}}}{\hbar}\mathcal{K}(\boldsymbol{r})\,\boldsymbol{\nabla}\times\boldsymbol{Q}(\boldsymbol{r})}_{\textrm{spin-current}}
	\label{ZORAtttcurrent}
\end{equation}
\end{widetext}
From the previous discussion, it is clear that in the absence of the vector potential (both the external one and the one induced by nuclear magnetic dipoles) the current density in 
Eq.~\eqref{ZORAtttcurrent} vanishes. A detailed derivation of the spin contributions is provided in Appendix~\ref{Section Spin Contributions to Current Density}, while the orbital contribution is taken directly from 
Ref.~\citenum{romaniello_relativistic_2007}. 

This derivation of the ZORA current density is in agreement with the expressions reported in Ref.~\citenum{romaniello_relativistic_2007}. In the limit of vanishing relativistic effects, where 
$\mathcal{K}(\boldsymbol{r}) \to \frac{1}{2m_{\textrm{e}}}$, the previous equation 
reduces to Eq.~\eqref{totalcurrent}, 
with the exception of the last singular term,
which appears in the Breit-Pauli formulation but not in the ZORA treatment. 
Also in this case the continuity equation, Eq.~\eqref{conteq},
is satisfied~\cite{romaniello_relativistic_2007,PhysRevLett.114.066404} for variational calculations in the limit of a complete basis set, see Appendix \ref{continuity} for a detailed discussion on spin contribution.  

So far, no further approximations have been made, apart from the choice of the ZORA Hamiltonian.
In the following discussion, however, relativistic corrections to the orbital current will be neglected, focusing instead on spin contributions. This choice is motivated by the observation that, in open-shell systems, spin-related terms generally represent the dominant contribution and exhibit high sensitivity to relativistic effects across the entire periodic table. 
In contrast, relativistic corrections to orbital contributions governing shielding and magnetizability tensors typically become significant only for heavier elements ($Z \geq 30 $)~\cite{speelman_nmr_2025}, as 
they do for
other magnetic properties such as indirect nuclear spin-spin coupling constants~\cite{Yuan2024}.

Thus, we will evaluate the orbital contributions using non-relativistic Hamiltonians, combined with the 
Continuous Transformation of the Origin of Current Density (CTOCD) 
approach~\cite{keith_calculation_1992,Keith_calculation_1993,monaco_atomic_2020,monaco_program_2021} to address the well-known issue of gauge-origin dependence. 
A detailed analysis of relativistic effects on the orbital current lies beyond the scope of the present work and is deferred to future studies. 
From a practical standpoint, this implies that, within our approximated treatment, both the spin Zeeman and spin–orbit coupling terms are omitted from the perturbative scheme. Instead, we employ the non-relativistic Hamiltonian as described by ~\citet{monaco_program_2021}.

Finally, in our treatment we neglect the picture-change effect. As shown and discussed in Refs.~\citenum{van_lenthe_relativistic_1994} and \citenum{romaniello_relativistic_2007}, the approximate ZORA density closely reproduces the one obtained with the Dirac(-Coulomb) Hamiltonian, particularly in the valence region, which is crucial for the evaluation of nuclear magnetic shielding tensors.

\subsection{Reduced Spin Density Formalism}

Consider the ground state of an open-shell molecule with total spin quantum number $S \neq 0$. This state consists of a $(2S+1)$-degenerate multiplet described by the eigenfunctions $\ket{S, M_S}$, where the spin projection quantum number spans $M_S = -S, -S+1, \dots, S$ along an arbitrary quantization axis $z$. In principle, the spin density 
$Q_{\gamma}(\boldsymbol{r})$ (with $\gamma$ denoting a component of the  vector) depends on the specific component of the multiplet under consideration. 
However, in the absence of SOC, the Wigner-Eckart theorem implies that spin densities are all the same except for a proportionality constant~\cite{mcweeny_spins_1970}. 

To exploit this symmetry, it is expedient to introduce an effective spin density operator $\mathcal{Q}_{\text{op},\gamma}(\boldsymbol{r})$ that is proportional to a reduced scalar function common to the entire multiplet \cite{soncini_charge_2007,mcweeny_spins_1970}:
\begin{equation}
	\mathcal{Q}_{\text{op},\gamma}(\boldsymbol{r}) = \frac{Q(\boldsymbol{r})}{S} \delta_{\gamma z} S_{\text{op},z} = Q_S(\boldsymbol{r}) \delta_{\gamma z} S_{\text{op},z}
	\label{eq:eff_spin_op}
\end{equation}
where:
\begin{itemize}
	\item $Q(\boldsymbol{r})$ is the spin density component along the quantization axis $z$ corresponding to the maximally polarized state $\ket{S, M_S = S}$. 
	\item $Q_S(\boldsymbol{r}) = Q(\boldsymbol{r})/S$ is the reduced spin density, a spatial function common to all components of the multiplet.
	\item $S_{\text{op},z}$ is the total spin projection operator along the $z$ axis.
\end{itemize}
The expectation value of the spin density for a given state $\ket{S, M_S}$ relates to the spin projection as follows:
\begin{equation}
	\int \bra{S, M_S} \mathcal{Q}_{\text{op},\gamma}(\boldsymbol{r}) \ket{S, M_S} d^3r = \langle S_{\text{op},\gamma} \rangle = M_S \delta_{\gamma z}
\end{equation}
The physical implications of this model are:
\begin{enumerate}
	\item Zero-Field Limit: In the absence of an external magnetic field, averaging Eq.~\eqref{eq:eff_spin_op} over all $2S+1$ degenerate components yields a vanishing net spin density.
	\item Magnetic Field Interaction: When a magnetic field is applied, the Zeeman interaction lifts the degeneracy of the multiplet. The field direction provides a physical quantization axis, inducing a non-zero spin density polarization.
	\item Singlet States: In the absence of spin-orbit coupling and hyperfine interactions, the spin density of a singlet state ($S=0$) is zero at every point in space. Consequently, singlet states do not contribute to the spin-dependent paramagnetic properties discussed herein.
\end{enumerate}

In the presence of non-vanishing SOC, the molecular magnetic response can no longer be treated within a conventional scalar or collinear framework. Instead, the spatial distribution of the three induced spin density components exhibits a dependence on the specific direction of the applied external magnetic field. The following discussion focuses exclusively on the calculation of the reduced spin density assuming a vanishingly small SOC interaction. Specific details are provided in the implementation section~\ref{Implementation}.

\subsection{Current Density Formulations}
\label{Current_Density_Formulations}

We will now focus on the 
derivation of orbital and spin contributions to nuclear magnetic shielding, nuclear hyperfine coupling and magnetizability tensors. 
The cornerstone of these derivations is the first-order current density vector for the spin multiplet, defined within a linear response framework as:
 \begin{equation}  
 	J_{\alpha}^{(1)}(\boldsymbol{r}) = \mathcal{J}_{\alpha}^{B_{\beta}}(\boldsymbol{r}) B_{\beta} + \mathcal{J}_{\alpha}^{S_{\beta}}(\boldsymbol{r}) S_{\text{op},\beta} 
 \end{equation}
 where the second-rank Current Density Tensors (CDTs) \cite{CDT} are defined as:
 \begin{equation}
 	\mathcal{J}_{\alpha}^{B_{\beta}}(\boldsymbol{r}) = \frac{\partial J_{\alpha}^{\boldsymbol{B}}(\boldsymbol{r})}{\partial B_{\beta}} \quad \text{and} \quad \mathcal{J}_{\alpha}^{S_{\beta}}(\boldsymbol{r}) = \frac{\partial J_{\alpha}^{\boldsymbol{S}}(\boldsymbol{r})}{\partial S_{\text{op},\beta}}~.
 	\label{CDT_defs}
 \end{equation}
 This approach provides a state-independent representation of the current density response, effectively decoupling the spatial distribution of the current from the specific spin projection $M_S$ of the multiplet.
 
\subsubsection{Nuclear Magnetic Shielding Tensor}

The total electronic energy of a molecule in the presence of external magnetic field $\boldsymbol{B}$ and intramolecular  magnetic dipoles $\boldsymbol{m}_{I}$, contains terms involving the NMR spectral parameters
\begin{equation}
W=W^{(0)}+\sigma_{\alpha\beta}^{I}m_{I\alpha}B_{\beta}+\cdots
\end{equation}
where the nuclear magnetic shielding at nucleus $I$ is defined as
\begin{equation}
	\sigma_{\alpha \beta}^I=\left.\frac{\partial^2 W}{\partial m_{I\alpha} \partial B_\beta}\right|_{\boldsymbol{m}_I, \boldsymbol{B} \rightarrow \boldsymbol{0}}
	\label{NMRshield}
\end{equation}
According to classical electrodynamics, an expression for the interaction energy between the current density vector and the vector potential given by the nuclear magnetic dipole $\boldsymbol{A}^{\boldsymbol{m}_{I}}$ can be obtained from 
the 
equation
\begin{equation}
W^{I}=-\int \,\boldsymbol{J}^{(1)}(\boldsymbol{r})\cdot\boldsymbol{A}^{\boldsymbol{m}_I}(\boldsymbol{r})\,d^3 r~.
	\label{classicalenergy}
\end{equation}
For a nucleus $I$ the interaction energy  between the induced total first order current density and the magnetic field generated by the nuclear magnetic dipole moment can be written as sum of two contributions, i.e., a spin independent and a spin dependent term
\begin{equation}
	W^{I}=-\int \,\left[\boldsymbol{J}^{\boldsymbol{B}}(\boldsymbol{r})+\boldsymbol{J}^{\boldsymbol{S}}(\boldsymbol{r})\right]\cdot\boldsymbol{A}^{\boldsymbol{m}_I}(\boldsymbol{r})\,d^3 r=W^{I\boldsymbol{B}}+W^{I\boldsymbol{S}}
\end{equation}
given by
\begin{equation}
	\begin{split}
	W^{I\boldsymbol{B}}=&-\int \mathcal{J}_{\lambda}^{B_{\beta}}(\boldsymbol{r})B_{\beta}\,A_{\lambda}^{m_I}(\boldsymbol{r})\,d^3 r=\\&-\frac{\mu_0}{4 \pi}\epsilon_{\lambda\alpha\gamma}\int \frac{m_{I\alpha}\left({r}_{\gamma}-{R}_{I\gamma}\right)}{\left|\boldsymbol{r}-\boldsymbol{R}_I\right|^3}\mathcal{J}_{\lambda}^{B_{\beta}}(\boldsymbol{r})B_{\beta}\,d^3 r
\end{split}
\end{equation}
for the spin-independent part, and
\begin{equation}
	\begin{split}
	W^{I\boldsymbol{S}}=&-\int \mathcal{J}_{\lambda}^{S_{\beta}}(\boldsymbol{r})S_{\text{op},\beta} A_{\lambda}^{m_I}(\boldsymbol{r})\,d^3 r=\\&-\frac{\mu_0}{4 \pi}\epsilon_{\lambda\alpha\gamma}\int \frac{m_{I\alpha}\left({r}_{\gamma}-{R}_{I\gamma}\right)}{\left|\boldsymbol{r}-\boldsymbol{R}_I\right|^3}\mathcal{J}_{\lambda}^{S_{\beta}}(\boldsymbol{r})S_{\text{op},\beta}\,d^3 r
\end{split}
\end{equation}
for the spin-dependent part. Using the expression in 
Eq.~\eqref{NMRshield}, it is clear that the spin independent term is
\begin{equation}
	\begin{split}
	\sigma_{\alpha \beta}^{I\boldsymbol{B}}
     =&\left.\frac{\partial^2 W^{I\boldsymbol{B}}}{\partial m_{I\alpha} \partial B_\beta}\right|_{\boldsymbol{m}_I, \boldsymbol{B} \rightarrow \boldsymbol{0}}\\
    =&-\frac{\mu_0}{4 \pi}\epsilon_{\lambda\alpha\gamma}\int \frac{{r}_{\gamma}-{R}_{I\gamma}}{\left|\boldsymbol{r}-\boldsymbol{R}_I\right|^3}\mathcal{J}_{\lambda}^{B_{\beta}}(\boldsymbol{r})\,d^3 r~.
\end{split}
	\label{orbshield}
\end{equation}
For the spin dependent term we can substitute $S_{\text{op},\beta}$ with its average value along the spin quantization axis for $g_{\textrm{e}}\mu_{B}|\boldsymbol{B}| \ll k_B T$~\cite{soncini_charge_2007,lewis2020thermodynamics}
\begin{equation}
	\langle S_{\text{op},\beta}\rangle=-g_{\textrm{e}}\mu_{B}B_{\beta}\frac{S(S+1)}{3k_{B}T}~,
\end{equation}
 where $k_B$ is the Boltzmann constant and $T$ is the temperature. A complete treatment of the spin statistics is provided in Appendix \ref{TERMOS}. Using this expression, we have
\begin{equation}
	\langle W^{I\boldsymbol{S}}\rangle =-\frac{\mu_0}{4 \pi}\langle S_{\text{op},\beta}\rangle\epsilon_{\lambda\alpha\gamma}\int \frac{m_{I\alpha}\left({r}_{\gamma}-{R}_{I\gamma}\right)}{\left|\boldsymbol{r}-\boldsymbol{R}_I\right|^3}\mathcal{J}_{\lambda}^{S_{\beta}}(\boldsymbol{r})\,d^3 r
	\label{spinenergy}
\end{equation}
from which it follows, within the Van-Vleck approximation~\cite{soncini_charge_2007}, that 
\begin{equation}
	\begin{split}
	\sigma_{\alpha \beta}^{I\boldsymbol{S}}=&\left.\frac{\partial^2 	\langle W^{I\boldsymbol{S}}\rangle }{\partial m_{I\alpha} \partial B_\beta}\right|_{\boldsymbol{m}_I, \boldsymbol{B} \rightarrow \boldsymbol{0}}\\=&\frac{\mu_0}{4 \pi}g_{\textrm{e}}\mu_{B}\frac{S(S+1)}{3k_{B}T}\epsilon_{\lambda\alpha\gamma}\int \frac{{r}_{\gamma}-{R}_{I\gamma}}{\left|\boldsymbol{r}-\boldsymbol{R}_I\right|^3}\mathcal{J}_{\lambda}^{S_{\beta}}(\boldsymbol{r})\,d^3 r
		\end{split}
	\label{spinshield}
\end{equation}
As emphasized in Eq.~\eqref{spinshield}, the resulting spin current density, when integrated according to the Biot-Savart law, provides a spatial mapping of the shielding contributions consistent with the shielding density concept developed by Jameson and Buckingham~\cite{jameson_nuclear_1979}. 
The dimensional analysis for the two terms \eqref{orbshield} and \eqref{spinshield} is as follows
\begin{eqnarray}
	\nonumber
	\textrm{dim}\left[\sigma_{\alpha \beta}^{I\boldsymbol{B}}\right]&=&\frac{\textrm{N}}{\textrm{A}^2}	\frac{1}{\textrm{m}^2}\frac{\textrm{A}}{\textrm{m}^2\textrm{T}}\textrm{m}^3=\frac{\textrm{N}}{\textrm{A}\,\textrm{m}\,\textrm{T}}=\frac{\textrm{N A m}}{\textrm{A m N}}\\
	\textrm{dim}\left[\sigma_{\alpha \beta}^{I\boldsymbol{S}}\right]&=&\frac{\textrm{N}}{\textrm{A}^2}\frac{1}{\textrm{T}}\frac{1}{\textrm{m}^2}\frac{\textrm{A}}{\textrm{m}^2}\textrm{m}^3=\frac{\textrm{N}}{\textrm{A}\,\textrm{m}\,\textrm{T}}=\frac{\textrm{N A m}}{\textrm{A m N}}
	\nonumber
\end{eqnarray} 
As can be seen, the nuclear magnetic shielding is a dimensionless quantity in SI units \cite{summa_molecular_2024}. 

In the scalar relativistic regime, the last term of the spin current \eqref{ZORAtttcurrent} is omitted due to the absence of spin-orbit coupling, and will therefore be neglected throughout the remainder of our discussion. The definition of the spin-contribution to chemical shift introduced here takes into account both contact and dipolar contributions~\cite{soncini_charge_2007,pennanen_density_2005,speelman_nmr_2025,blugel_hyperfine_1987}. 
For isotropically tumbling molecules, the (intrinsically traceless) spin dipolar term contributes to the chemical shift only through the anisotropic part of the paramagnetic susceptibility tensor — which vanishes, to a good approximation, in the absence of spin-orbit coupling.

If the g tensor also has an anisotropic component then the anisotropic dipolar part of the hyperfine coupling tensor matrix can also contribute to the isotropic chemical shift to give what is called the pseudo-contact term \cite{rinkevicius_calculations_2003,autschbach_chapter_2009}. 
Being in the scalar relativistic regime we will not take into account this effect, the anisotropic component of the g tensor being given by spin Zeeman and spin-orbit coupling interactions.

\subsubsection{Nuclear Hyperfine Coupling Tensor}

Using an approach similar to the one above we can compute nuclear hyperfine coupling (HFC) tensors as 
\begin{equation}
	A_{\alpha \beta}^I=\frac{1}{h}\left.\frac{\partial^2 W^{I\boldsymbol{S}}}{\partial I_{I\alpha} \partial S_\beta}\right|_{\boldsymbol{I}_I, \boldsymbol{S} \rightarrow \boldsymbol{0}}
\end{equation}
Using Eq.~\eqref{spinenergy}, it follows that
\begin{equation}
	A_{\alpha \beta}^I=-\frac{g_I\mu_{N}}{h}\frac{\mu_0}{4\pi}\epsilon_{\lambda\alpha\gamma}\int \frac{{r}_{\gamma}-{R}_{I\gamma}}{\left|\boldsymbol{r}-\boldsymbol{R}_I\right|^3}\mathcal{J}_{\lambda}^{S_{\beta}}(\boldsymbol{r})\,d^3 r~.
	\label{hyperfineA}
\end{equation}
Its dimensional analysis yields 
\begin{equation}
	\nonumber 
	\textrm{dim}\left[A_{\alpha \beta}^{I}\right]=\frac{1}{\textrm{T\,s}}\frac{\textrm{N}}{\textrm{A}^2}\frac{1}{\textrm{m}^2}\frac{\textrm{A}}{\textrm{m}^2}\textrm{m}^3=\frac{1}{\textrm{s}}=\textrm{Hz}
\end{equation}
so units of $A^I_{\alpha\beta}$ are $\left[\textrm{Hz}\right]$ in the SI system. 
As can be seen, by inspecting Eqs. (\ref{spinshield}) and (\ref{hyperfineA}), nuclear hyperfine coupling tensors and the spin contribution to nuclear magnetic shielding tensors are mathematically intertwined, as discussed in Refs. \citenum{soncini_charge_2007} and  \citenum{franzke_paramagnetic_2024}.

\subsubsection{Magnetizability Tensor}

The total electronic energy of a molecule in the presence of external magnetic field $\boldsymbol{B}$ can be written as
\begin{equation}
	W=W^{(0)}-\frac{1}{2}\chi_{\alpha\beta}B_{\alpha}B_{\beta}+\cdots
\end{equation}
where the susceptibility tensor $\chi_{\alpha\beta}$ is given by
\begin{equation}
	\chi_{\alpha \beta}=-\left.\frac{\partial^2 W}{\partial B_{\alpha} \partial B_\beta}\right|_{\boldsymbol{B} \rightarrow \mathbf{0}}
	\label{magnetizability}
\end{equation}
According to classical electrodynamics, an expression for the interaction energy between the current density vector and the vector potential $\boldsymbol{A}^{\boldsymbol{B}}$ at second order in perturbation theory is used \cite{lazzeretti_methods_2012,Lazzeretti2012erratum,kern_magnetic_1962,stevens_perturbed_1963}
\begin{equation}
	W^{\boldsymbol{B}}=-\frac{1}{2}\int \,\boldsymbol{J}^{(1)}(\boldsymbol{r})\cdot\boldsymbol{A}^{\boldsymbol{B}}(\boldsymbol{r})\,d^3 r
	\label{classicalenergymagn}
\end{equation}
Also in this case 
Eq.~\eqref{classicalenergymagn} can be written as sum of two contributions, i.e., a 
spin-independent and a 
spin-dependent term
\begin{equation}
	W^{\boldsymbol{B}}=-\frac{1}{2}\int \left[\boldsymbol{J}^{\boldsymbol{B}}(\boldsymbol{r})+\boldsymbol{J}^{\boldsymbol{S}}(\boldsymbol{r})\right]\cdot\boldsymbol{A}^{\boldsymbol{B}}(\boldsymbol{r})\,d^3 r=W^{\boldsymbol{B}\boldsymbol{B}}+W^{\boldsymbol{B}\boldsymbol{S}}
\end{equation}
with
\begin{equation}
	\begin{split}
	W^{\boldsymbol{BB}}=&-\int \mathcal{J}_{\alpha}^{B_{\delta}}(\boldsymbol{r})B_{\delta}\,A_{\alpha}^{B}(\boldsymbol{r})\,d^3 r=\\&-\frac{1}{4}\epsilon_{\alpha\beta\gamma}B_{\delta}B_{\beta}\int r_{\gamma}\,\mathcal{J}_{\alpha}^{B_{\delta}}(\boldsymbol{r})\,d^3 r
\end{split}
\end{equation}
and
\begin{equation}
		\begin{split}
	W^{\boldsymbol{BS}}=&-\int \mathcal{J}_{\alpha}^{S_{\delta}}(\boldsymbol{r})S_{\text{op},\delta} A_{\alpha}^{B}(\boldsymbol{r})\,d^3 r=\\&-\frac{1}{4}\epsilon_{\alpha\beta\gamma}S_{\text{op},\delta}B_{\beta}\int r_{\gamma}\,\mathcal{J}_{\alpha}^{S_{\delta}}(\boldsymbol{r}) \,d^3 r~.
\end{split}
\end{equation}
Using Eq.~\eqref{magnetizability}, it is clear that the 
spin-independent term is~\cite{summa_molecular_2024}
\begin{equation}
	\begin{split}
		\chi_{\mu \lambda} ^{\boldsymbol{B}}&=-\left.\frac{\partial^2 W^{\boldsymbol{BB}}}{\partial B_{\mu} \partial B_\lambda}\right|_{\boldsymbol{B} \rightarrow \mathbf{0}}\\
		& =\frac{1}{4}  \int (\epsilon_{\lambda \gamma \alpha}\mathcal{J}_\alpha^{B_\mu}+\epsilon_{\mu \gamma \alpha}\mathcal{J}_\alpha^{B_\lambda}) r_\gamma \,d^3 r
	\end{split}
	\label{orbmagn}
\end{equation}
As done before for the spin dependent term of the shielding, 
we can substitute $S_{\text{op},\delta}$ with its average value along the spin 
quantization axis~\cite{soncini_charge_2007} and obtain~\cite{lazzeretti_methods_2012,Lazzeretti2012erratum} 
\begin{equation}
	\begin{split}
		\chi_{\mu \lambda} ^{\boldsymbol{S}}&=-\left.\frac{\partial^2 \langle W^{\boldsymbol{BS}}\rangle}{\partial B_{\mu} \partial B_\lambda}\right|_{\boldsymbol{B} \rightarrow \mathbf{0}} \\
		& =-g_{\textrm{e}}\mu_{B}\frac{S(S+1)}{12k_{B}T} \int (\epsilon_{\lambda \gamma \alpha}\mathcal{J}_\alpha^{S_\mu}+\epsilon_{\mu \gamma \alpha}\mathcal{J}_\alpha^{S_\lambda}) r_\gamma \,d^3 r
	\end{split}
	\label{spinmagn}
\end{equation}
Also in this case a dimensional analysis can be performed for both terms (\ref{orbmagn}) and (\ref{spinmagn}), yielding
\begin{equation}
	\nonumber
	\textrm{dim}\left[\chi_{\alpha \beta}^{\boldsymbol{B}}\right]=\textrm{m}\frac{\textrm{A}}{\textrm{m}^2\,\textrm{T}}\textrm{m}^3=\frac{\textrm{A}\,\textrm{m}^2}{\textrm{T}}=\frac{\textrm{J}}{\textrm{T}^2}
\end{equation}
\begin{equation}
	\textrm{dim}\left[\chi_{\alpha \beta}^{\boldsymbol{S}}\right]=\frac{1}{\textrm{T}}\textrm{m}\frac{\textrm{A}}{\textrm{m}^2}\textrm{m}^3=\frac{\textrm{A}\,\textrm{m}^2}{\textrm{T}}=\frac{\textrm{J}}{\textrm{T}^2}
	\nonumber
\end{equation}
so 
units of $\chi$ in the SI system  are $\left[\textrm{J}\textrm{T}^{-2}\right]$ \cite{summa_molecular_2024}.

\section{Implementation at GHF-GKS or HF-DFT levels of Theory} 
\label{Implementation}

\subsection{Background}
In Generalized Hartree-Fock (GHF) or Generalized Kohn-Sham (GKS),
the wavefunction $\Psi$ is represented by a single Slater determinant constructed from $n$ occupied two-component molecular spinors $\psi_{i}(\boldsymbol{r})$ as:
\begin{equation}
	\Psi = \frac{1}{\sqrt{n!}} \textrm{det} \left[ \psi_1, \psi_2, \dots, \psi_n \right]
\end{equation}
When SOC is considered, the symmetry between different spinor components is broken, and the spatial part of these spinors must be expanded as linear combinations of basis functions $\chi_q(\boldsymbol{r})$. In this context, each occupied spinor $\psi_i(\boldsymbol{r})$ incorporates both $\alpha$ and $\beta$ components, expanded as:
\begin{equation}
	\psi_i(\boldsymbol{r})=\sum_{q}\left[c_{qi}^\alpha\left(\begin{array}{c}
		\chi_q(\boldsymbol{r}) \\
		0
	\end{array}\right)+c_{qi}^\beta\left(\begin{array}{c}
		0 \\
		\chi_q(\boldsymbol{r})
	\end{array}\right)\right]
\end{equation}
where the coefficients $c_{qi}^\gamma$, with $\gamma = \alpha,\beta$ are generally complex to account for the non-collinear nature of the electronic system:
\begin{equation}
	c_{qi}^\gamma = c_{qi}^{\Re,\gamma} + \mathrm{i}c_{qi}^{\Im,\gamma}
\end{equation}
This leads to the explicit form of the $i$-th occupied spinor and its corresponding adjoint
\begin{align}
	&\psi_i(\boldsymbol{r})=\sum_{q}\left[\begin{array}{c}
		(c_{qi}^{\Re,\alpha} + \mathrm{i}c_{qi}^{\Im,\alpha})\chi_q(\boldsymbol{r}) \\
		(c_{qi}^{\Re,\beta} + \mathrm{i}c_{qi}^{\Im,\beta})\chi_q(\boldsymbol{r})
	\end{array}\right]
\\
	&\psi_i^{\dagger}(\boldsymbol{r})=\sum_{p}\left[\begin{array}{c}
		(c_{pi}^{\Re,\alpha} - \mathrm{i}c_{pi}^{\Im,\alpha})\chi_p(\boldsymbol{r}) \\
		(c_{pi}^{\Re,\beta} - \mathrm{i}c_{pi}^{\Im,\beta})\chi_p(\boldsymbol{r})
	\end{array}\right]^{T}
\end{align}
where $T$ means vector transposition. From these definitions, the probability charge density $\gamma(\boldsymbol{r})$ and the spin density vector components $Q_\alpha(\boldsymbol{r})$ can be explicitly derived by using their definition as follows:
\begin{align}
	&\gamma(\boldsymbol{r})=\sum_i^{occ}\psi_i^{\dagger}(\boldsymbol{r})\psi_i(\boldsymbol{r})
\\
	&Q_{\alpha}(\boldsymbol{r})=\frac{\hbar}{2}\sum_{i}^{occ}\psi_i^{\dagger}(\boldsymbol{r})\hat{\sigma}_{\alpha}\psi_i(\boldsymbol{r})	\qquad \alpha=x,y,z
\end{align}
By using these expressions, we can define the probability charge density and the spin density vector in terms of density matrices as
\begin{align}
&\gamma(\boldsymbol{r})=\sum_{pq}P_{pq}\chi_p(\boldsymbol{r})\chi_q(\boldsymbol{r})
    \label{gamma_ZORA}\\
	&Q_{\alpha}(\boldsymbol{r})=\frac{\hbar}{2}\sum_{pq}P_{pq}^{\sigma_{\alpha}}\chi_p(\boldsymbol{r})\chi_q(\boldsymbol{r}) \qquad\alpha=x,y,z
    \label{q_ZORA}
\end{align}
where
\begin{align}
	&P_{pq}=\sum_{i}^{occ}\left[c_{pi}^{\Re,\alpha}c_{qi}^{\Re,\alpha}+c_{pi}^{\Re,\beta}c_{q i}^{\Re,\beta}+c_{pi}^{\Im,\alpha}c_{qi}^{\Im,\alpha}+c_{pi}^{\Im,\beta}c_{qi}^{\Im,\beta}\right]\\
	&P_{pq}^{\sigma_x}=\sum_{i}^{occ}
	\left[c_{pi}^{\Re,\alpha}c_{qi}^{\Re,\beta}+c_{pi}^{\Re,\beta}c_{qi}^{\Re,\alpha}+c_{pi}^{\Im,\alpha}c_{qi}^{\Im,\beta}+c_{pi}^{\Im,\beta}c_{qi}^{\Im,\alpha}\right]\\
	&P_{pq}^{\sigma_y}=\sum_{i}^{occ}\left[c_{pi}^{\Re,\alpha}c_{qi}^{\Im,\beta}+c_{pi}^{\Im,\beta}c_{qi}^{\Re,\alpha}-c_{pi}^{\Im,\alpha}c_{qi}^{\Re,\beta}-c_{pi}^{\Re,\beta}c_{qi}^{\Im,\alpha}\right]\\
	&P_{pq}^{\sigma_z}=\sum_{i}^{occ}\left[c_{pi}^{\Re,\alpha}c_{qi}^{\Re,\alpha}-c_{pi}^{\Re,\beta}c_{qi}^{\Re,\beta}+c_{pi}^{\Im,\alpha}c_{qi}^{\Im,\alpha}-c_{pi}^{\Im,\beta}c_{qi}^{\Im,\beta}\right]
\end{align} 
The previously derived Eqs.~\eqref{gamma_ZORA} and 
\eqref{q_ZORA} can be applied in a general finite field scheme and used to obtain both unperturbed ground state and perturbed densities. 

As can be seen for a real unrestricted Hartree–Fock (UHF) or Kohn–Sham (UKS) approach, in the absence of SOC and magnetic interactions, no mixing occurs between the $\alpha$ and $\beta$ components. Consequently, for open-shell molecules, only the $Q_z$ spin density survives. This follows from symmetry considerations, as the imaginary parts of the coefficients, \( c_{qi}^{\Im,\gamma} \), vanish.
If we start from an unrestricted calculation (UHF or UKS) as a guess, we can choose $z$ as the quantization axis and compute the isotropic reduced spin density according to 
Eq.~\eqref{eq:eff_spin_op} as
\begin{equation}
Q_S(\boldsymbol{r})=\frac{Q_z^{(0)}}{S}
\end{equation}
This approach is exact for scalar relativistic corrections only, in the limit of a vanishing magnetic field, but not when SOC is accounted for. 

\subsection{Approximate Treatment of SOC Contributions}

 At present, a full Coupled-Perturbed Generalized Hartree-Fock (CP-GHF) or Coupled-Perturbed Generalized Kohn-Sham (CP-GKS) scheme incorporating both Spin Zeeman and field-dependent SOC contributions within ZORA has not yet been implemented in SYSMOIC~\cite{monaco_program_2021}.  This is because the program currently lacks relativistic GHF/GKS code under a magnetic field, which is required to extract the linear response using a GHF/GKS-level perturbation theory that accounts for both SOC and magnetic interactions.
 
 We can nevertheless introduce a simplified, twofold approximation scheme designed specifically as a diagnostic benchmark. By systematically decoupling the different relativistic response sources, this framework allows us to evaluate the performance of a \textit{field-free SOC} approach against standard non-relativistic treatments, thereby identifying the exact thresholds where a rigorous, field-dependent CP-GKS/CP-GHF implementation becomes physically mandatory.

In this testing framework, we introduce two main constraints: first, while the orbital response to the external magnetic field is fully included, it is treated at the standard non-relativistic level. Second, we neglect the explicit field-dependence of the SOC in the linear response, evaluating the spin density modifications solely through the lens of the field-free ground-state anisotropy. This strategy yields significant computational savings and is expected to hold for lighter systems up to the 3d transition metal series ($Z \simeq 30$). More importantly, it provides a minimal ansatz to map out precisely where non-collinear ground-state effects suffice and where explicit field-dependent SOC contributions can no longer be ignored.

In a rigorous linear response treatment, the components of the spin density vector $\boldsymbol{Q}(\boldsymbol{r})$ evolve under an applied static magnetic field $B_{\gamma}$ as:
\begin{equation}
Q_{\alpha}(\boldsymbol{r}) = Q_{\alpha}^{(0)}(\boldsymbol{r}) + Q_{\alpha}^{B_{\gamma}}(\boldsymbol{r})B_{\gamma}
\end{equation}
where $Q_{\alpha}^{(0)}(\boldsymbol{r})$ is the ground-state (field-free) spin density component, and $Q_{\alpha}^{B_{\gamma}}(\boldsymbol{r})$ is the linear response tensor driven by $\hat{h}_k^{(1)}$. 

Within our framework, $\hat{h}_k^{(1)}$ is evaluated by approximating the ZORA factor $\mathcal{K}(\boldsymbol{r})$ as $\frac{1}{2m_{\textrm{e}}}$ and by removing the other two spin terms, thereby reducing it to its standard non-relativistic form 
\begin{equation}
    \frac{e}{2m_{\textrm{e}}}\boldsymbol{\hat{L}}\cdot\boldsymbol{B}
\end{equation}
Furthermore, the core of our spin-response approximation lies in the complete neglect of the explicit field-dependent linear response term, $Q_{\alpha}^{B_{\gamma}}(\boldsymbol{r}) \simeq 0$. The magnetic response of the spin density is thus modulated exclusively by the field-free ground-state wavefunction (hence the term \textit{field-free SOC response approximation}).

To understand how the system's anisotropy is modeled within this framework, it is instructive to look at the non-relativistic or scalar-relativistic limit (vanishing SOC). In this regime, the spin and spatial degrees of freedom are decoupled, making the spin space perfectly isotropic. Under an external field, the induced spin density aligns collinearly with the field direction. In the limit of a vanishingly small field ($\boldsymbol{B} \to 0$), the response along any arbitrary spatial axis $x, y,$ or $z$ is identical and maps onto the unperturbed longitudinal spin density, $Q_{z}^{(0)}(\boldsymbol{r})$:
\begin{equation}
Q_x(\boldsymbol{r}) = Q_y(\boldsymbol{r}) = Q_z(\boldsymbol{r}) = Q_{z}^{(0)}(\boldsymbol{r})
\end{equation}
where $z$ is chosen as the quantization axis for the field-free state.

When \textit{field-free SOC response} is restored, it introduces a spatial anisotropy into the unperturbed components $Q_{\alpha}^{(0)}(\boldsymbol{r})$. Following the formalism of~\citet{soncini_charge_2007}, we account for these relativistic non-collinear effects by replacing the standard longitudinal magnetization with the total effective spin magnitude $S$:
\begin{equation}
    S = \sqrt{\sum_{\alpha} \left\{\int Q_{\alpha}^{(0)}(\boldsymbol{r})\,d^3r\right\}^2} 
\end{equation}
To satisfy the isotropic boundary condition required in the zero SOC limit, while simultaneously incorporating the relativistic anisotropy of the ground state, we define the \textit{directional reduced spin densities} $Q_{S}^{\alpha}(\boldsymbol{r})$ for each field orientation as:
\begin{align}
&Q_{S}^{x}(\boldsymbol{r}) \simeq \frac{Q_{z}^{(0)}(\boldsymbol{r}) + Q_{x}^{(0)}(\boldsymbol{r})}{S} \\
&Q_{S}^{y}(\boldsymbol{r}) \simeq \frac{Q_{z}^{(0)}(\boldsymbol{r}) + Q_{y}^{(0)}(\boldsymbol{r})}{S} \\
&Q_{S}^{z}(\boldsymbol{r}) \simeq \frac{Q_{z}^{(0)}(\boldsymbol{r})}{S}
\end{align}
Here, the longitudinal term $Q_{z}^{(0)}(\boldsymbol{r})$ contains both the dominant collinear contribution and a relativistic correction due to spinor mixing. Conversely, the transverse terms $Q_{x}^{(0)}(\boldsymbol{r})$ and $Q_{y}^{(0)}(\boldsymbol{r})$ arise exclusively from the field-free spin-orbit interaction. Together, these three components describe a non-collinear ground-state spin distribution that deviates from the isotropic scalar-relativistic density. This minimal ansatz allows us  to evaluate when the inclusion of SOC induces local non-collinear deviations that can no longer be accommodated by a standard collinear model, thereby establishing the rigorous boundaries beyond which a full, explicit CP-GKS/CP-GHF treatment becomes mandatory.

\subsection{Implementation details}

To numerically evaluate the influence of both scalar and spin-orbit relativistic effects according to the diagnostic framework derived above, GKS calculations were performed in {Gaussian 16} \cite{g16}. It is worth emphasizing that these electronic structure calculations are used exclusively to obtain the unperturbed, field-free ground-state wavefunctions and densities, from which the zeroth-order components $Q_{\alpha}^{(0)}(\boldsymbol{r})$ are subsequently extracted.

Scalar relativistic effects were introduced via the \texttt{int=dkh} keyword, which requests a second-order DKH (DKH2) scalar relativistic calculation. This method effectively treats mass-velocity and Darwin terms and utilizes a Gaussian nuclear model to represent the finite nucleus, thereby avoiding the singularities associated with a point-charge potential. 

To evaluate our \textit{field-free SOC response} ansatz, the unperturbed spin-orbit coupling effects in the ground state were explicitly included using the \texttt{int=dkhso} option, which requests a fourth-order DKH (DKH4) relativistic calculation. This higher-order expansion is necessary to consistently incorporate the one-electron spin-orbit terms during the SCF procedure. This enables the mixing of spinor components already at the field-free level, thereby generating the non-zero transverse ground-state densities $Q_x^{(0)}(\boldsymbol{r})$ and $Q_y^{(0)}(\boldsymbol{r})$ that drive the spatial anisotropy in our directional model.

To maintain consistency with the ZORA formalisms implemented in SYSMOIC, a dedicated interface was developed to extract these electronic data from formatted checkpoint files (\texttt{.fchk}), generated with Cartesian basis functions (\texttt{6d 10f} keyword). Although DKH \cite{nakajima_douglaskrollhess_2012} is the only relativistic method available in Gaussian 16, it yields field-free ground-state densities virtually identical to those obtained via a ZORA approach \cite{hong_comparison_2001}, as both methods serve as systematic, highly convergent approximations to the Foldy-Wouthuysen transformation. This numerical equivalence ensures that the DKH-based ground-state densities extracted from Gaussian 16 are fully compatible with our ZORA-based diagnostic framework.

The ZORA approach has a significant advantage over DKH with respect to obtaining a closed analytical expression of the magnetically induced current density, since the latter is not easily obtained by using the Landau procedure. By employing the reduced spin density $Q_S(\boldsymbol{r})$, or $Q_S^{\alpha}(\boldsymbol{r})$ in 
Eqs.~\eqref{spinshield}, \eqref{hyperfineA} and \eqref{spinmagn}, we ensure consistency with this current density formulation. 

Note that in our approach we always assume that $N_{\alpha} > N_{\beta}$ where $N_{\alpha}$ and $N_{\beta}$ are the number of $\alpha$ and $\beta$ electrons, respectively, to have consistency with our previous reasoning. 

Numerical integration was carried out using the Becke algorithm with the Treutler-Alrichs variant~\cite{becke_numerical_1988,treutler_efficient_1995}, 
employing a radial mapping 
\begin{equation}
	r=\frac{r_m}{\ln(2)}\ln\left(\frac{2}{1-x}\right)
\end{equation}
where the total mean spherical radii $r_m$ for each atom were sourced from 
Ref.~[\citenum{luo_theoretical_2021}], with the exception of the H atom ($r_m=0.35$ a.u.) following Becke's recommendation \cite{becke_multicenter_1988}.
The atomic grid of each atom $i$ with atomic number $Z_i$ is constructed as a product of $N_{\text{rad}}$ radial and $N_{\text{ang}}$ angular grid points. This yields a total of $N_{\text{total}} = N_{\text{rad}} \times N_{\text{ang}}$ integration points per atom. The number of local radial points $N_{\text{rad}}$ is defined as a function of the user-selected grid-quality parameter $S_{\text{rad}}$ (\texttt{RADSIZE}) and the nuclear charge $Z_i$:
\begin{equation}
    N_{\text{rad}} = 40 + 10(S_{\text{rad}} - 1) + 2 \cdot \text{NINT}(Z_i)
\end{equation}
where $\text{NINT}$ denotes the nearest integer function. The parameter $S_{\text{rad}}$ maps to predefined quality levels:
\begin{equation}
    S_{\text{rad}} = \begin{cases} 
    1 & \text{\texttt{LOW}} \\
    2 & \text{\texttt{COARSE}} \\
    3 & \text{\texttt{MEDIUM}} \\
    4 & \text{\texttt{FINE}} \\
    5 & \text{\texttt{ULTRAFINE}} \\
    6 & \text{\texttt{SUPERFINE}}
    \end{cases}
\end{equation}
The local angular point count $N_{\text{ang}}$ is assigned dynamically based on the atomic number range $Z_i$ and the quality selection $S_{\text{rad}}$, as outlined in Table~\ref{tab:angular_grid_defaults}. 
\begin{table*}[htbp]
\centering
\caption{Default angular integration points ($N_{\text{ang}}$) as a function of grid quality and atomic charge ($Z_i$).}
\label{tab:angular_grid_defaults}
\begin{tabular}{lcccccc}
\hline\hline
Quality Level & $S_{\text{rad}}$ & $Z \le 2$ & $2 < Z \le 10$ & $10 < Z \le 36$ & $36 < Z \le 86$ & $86 < Z \le 118$ \\
\hline
\texttt{LOW}       & 1   & 110 & 194 & 302  & 434  & 590  \\
\texttt{COARSE}    & 2   & 194 & 302 & 434  & 590  & 770  \\
\texttt{MEDIUM}    & 3   & 302 & 434 & 590  & 770  & 974  \\
\texttt{FINE}      & 4   & 434 & 590 & 770  & 974  & 1202 \\
\texttt{ULTRAFINE} & 5   & 590 & 770 & 974  & 1202 & 1454 \\
\texttt{SUPERFINE} & 6   & 770 & 974 & 1202 & 1454 & 1730 \\
\hline\hline
\end{tabular}
\end{table*}
Angular distribution points and weights are evaluated via Lebedev quadrature. For all numerical calculations here reported, medium grid-quality level has been chosen.
The developed interface works up to g-type of Cartesian basis set functions and it is not able in the present implementation to deal with combined sp shells. The implementation of spin current density tensors was performed by extending the formalisms of 
Refs.~[\citenum{soncini_charge_2007,summa_molecular_2024}] through the introduction of the ZORA scaling factor and the effective potential described in Appendix \ref{potnuc}. This framework has been integrated as a new feature into the SYSMOIC software package \cite{monaco_program_2021}. 

Finally, for the orbital contributions to nuclear magnetic shielding and magnetizability, the CTOCD approach (using the 
Continuous Set of Gauge Transformations (CSGT) definition of the shift function) was employed to ensure gauge-origin independence.\cite{Keith_calculation_1993,keith_topological_1993,cheeseman_comparison_1996,monaco_atomic_2020,monaco_program_2021} It should be noted that, in the present implementation, the orbital contribution within the CTOCD framework is evaluated at the non-relativistic level. However, for the open-shell systems investigated in this work, the spin-dependent terms—where relativistic effects are most prominent—are expected to be significantly larger than the orbital contributions. Therefore, the leading relativistic corrections to the magnetic properties are effectively captured through the non-collinear spin density formalism described above. 

The magnetically induced current density maps depicted in Figures~\ref{SCUR2} and~\ref{SCUR} have been computed using the spin density of the maximum polarized state obtained from Gaussian 16~\cite{g16}.

\section{Results and Discussion}
\label{Results_and_Discussion}

A set of representative molecules has been selected to illustrate the application of the theory presented above for the calculation of nuclear magnetic shielding, nuclear hyperfine coupling constants and magnetizability tensors.

\subsection{Nuclear Magnetic Shieldings}

In the evaluation of the isotropic  nuclear magnetic shielding, the molecular dataset was selected following the work of \citet{pennanen_density_2005} and \citet{franzke_paramagnetic_2024} focusing on open-shell systems with varying spin multiplicities, particularly triplet states ($S=1$) and higher-spin transition metal complexes. These systems were chosen to evaluate the performance of both (\ref{totalcurrent}), with the absence of the last singular term, and (\ref{ZORAtttcurrent}) current densities. The molecular species, their spin multiplicities, the level of theory, the source of geometries, the considered temperature for the spin contribution and the solvent specifications, where available, are summarized in Table \ref{shmolecules}. Following~\citet{franzke_paramagnetic_2024}, eclipsed structures were considered for all metallocenes. 

\begin{table*}[hbpt!]
		\centering
	\caption{Molecular dataset specifications for chemical shift calculations: spin multiplicities ($2S+1$), level of theory, used temperature in kelvin, geometry references and solvent model.}
	\label{shmolecules}
	\begin{tabular}{lccccc}
		\hline
		Molecule & $2S+1$ & Level of Theory & T (K)&Geometry & Solvent \\ 
		\hline
		\ch{V[C5H5]2}&4  &B3LYP/X2C-QZVPall-s  &298.00&Ref. \citenum{franzke_paramagnetic_2024} \, &CPCM,toluene\\		
		\ch{Cr[C5H5]2}&3  &B3LYP/X2C-QZVPall-s  &298.00&Ref. \citenum{franzke_paramagnetic_2024} \, &CPCM,toluene\\
		\ch{Mn[C5H5]2}&6  &B3LYP/X2C-QZVPall-s  &390.00&Ref. \citenum{franzke_paramagnetic_2024} \, &CPCM,toluene\\        \ch{Co[C5H5]2}&2  &B3LYP/X2C-QZVPall-s  &298.00&Ref. \citenum{franzke_paramagnetic_2024} \,  &CPCM,toluene\\
		\ch{Ni[C5H5]2}&3  &B3LYP/X2C-QZVPall-s  &298.00&Ref. \citenum{franzke_paramagnetic_2024} \, &CPCM,toluene\\
		\ch{Rh[C5H5]2}&2  &B3LYP/X2C-QZVPall-s  &298.00&Ref. \citenum{rouf_relativistic_2017}  &CPCM,toluene\\
		\ch{C8H15N2O2}&2  &B3LYP/X2C-QZVPall-s  &298.00&MP2/6-31G(d)  &None\\
		\hline
	\end{tabular}
\end{table*}

\begin{figure}[htbp!]
    \centering
    \includegraphics[width=0.8\linewidth]{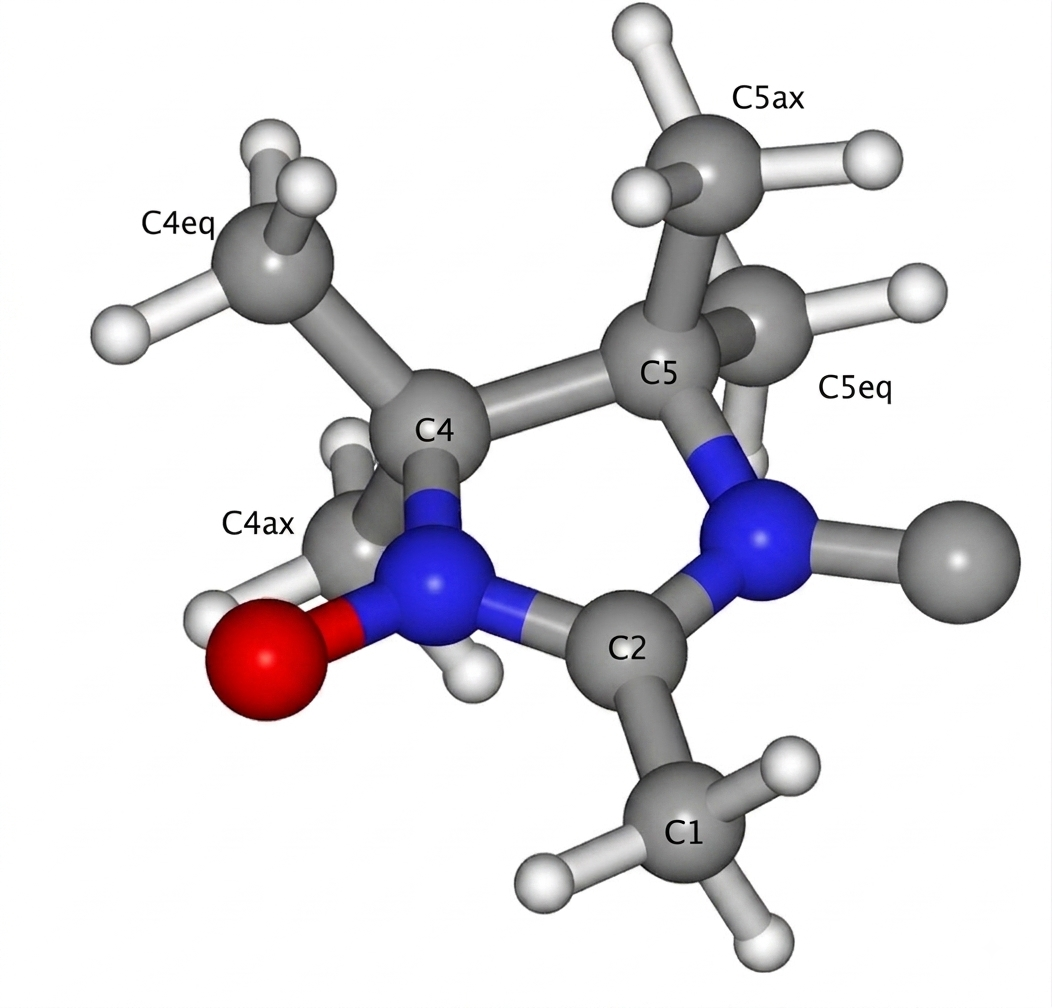}
    \caption{Nitroxide compound N6.}
    \label{N6Fig}
\end{figure}

In Table~\ref{metalloceni}  we report calculated $^1$H and $^{13}$C pNMR shieldings ($\sigma$) and chemical shifts ($\delta$), obtained using the shieldings from ferrocene as reference, for five 3d and one 4d metallocenes. Similar results are given in Table~\ref{N6} 
for the  N6 nitroxide radical of Ref.~[\citenum{pennanen_density_2005}]  (\ch{C8H15N2O2}). The 
labels used in Table \ref{N6} are shown in  Figure~\ref{N6Fig}. 

By comparing the shielding values obtained from the non-relativistic (NR) electron current density (Eq.~\eqref{totalcurrent}), $\sigma^\text{NR}_\text{iso}$, to the ones obtained with the ZORA spin current (Eq.~\eqref{ZORAtttcurrent}) considering only scalar relativistic contributions, $\sigma^\text{S(Z)}_\text{iso}$, we observe that 
the scalar relativistic effects on the $^1$H pNMR shieldings remain quite similar across the studied series of 3d and 4d metallocenes and of the order of 3-7 ppm. For the $^{13}$C pNMR shieldings, we typically observe larger differences that can reach 14-22 ppm. We note however that for the Mn and Co systems, the scalar relativistic effects on the $^1$H and $^{13}$C pNMR shieldings are of similar magnitude, and smaller than for the other species.

The same comparison between non-relativistic and scalar relativistic shieldings for \ch{C8H15N2O2} shows essentially no difference for the $^1$H pNMR shieldings, and effects of the order of a few ppm for the $^{13}$C pNMR shieldings (with the exception of about 25 ppm for C$_2$), and illustrates the importance of the scalar Heavy Atom on Light Atom (HALA) effect for the case of the metallocenes.

Comparing now the scalar relativistic pNMR shieldings, $\sigma^\text{S(Z)}_\text{iso}$, to those including the (approximate) treatment of SOC, $\sigma^\text{S(ZS)}_\text{iso}$, for the 3d metallocenes, we observe that, for both atoms, the change due to SOC is an order of magnitude smaller than that originating from scalar relativistic effects (though for the Mn system the change in $^{13}$C pNMR shielding is similar in magnitude but opposite in sign to the scalar relativistic contribution). As expected, for the 4d metallocene the spin-orbit contributions are more important than for the 3d systems, and in the case of $^{13}$C, larger than the scalar relativistic effect. 

The $^1$H nuclear magnetic shielding densities of metallocenes, defined by means of Eqs.~\eqref{orbshield} and \eqref{spinshield}, are shown in Figure~\ref{shielddens}. As can be seen, spin contributions are consistently dominant, accounting for nearly the entire total nuclear magnetic shielding in all investigated systems, as further illustrated by the numerical integrations of the respective densities reported in Table \ref{metalloceni}.

A comparison to experimental chemical shifts is possible for some of the 3d metallocenes and the \ch{C8H15N2O2} molecule. For the latter, most shifts differ from experiment by about 5-10\%, though in some cases significantly more. For the metallocenes, we see differences of about 7-20\%, though for the case of the Mn complex these are significantly higher.

Finally, we recall that our treatment of SOC foregoes the explicit solution of the 
coupled-perturbed Generalized Hartree-Fock/Kohn-Sham equations. While this approach 
admittedly lacks formal rigor, 
it is designed specifically as a diagnostic benchmark 
to map the regions where explicit spin-orbit response contributions become non-negligible; 
this makes it computationally efficient, but at the same time unable to fully account 
for the first-order changes to spin-densities arising from the spin-dependent part of 
the shielding tensors. The loss of accuracy of our approximation is admittedly 
unacceptable for Rhodocene; however, rather than a failure of the method, this 
discrepancy demonstrates the success of our approach as a diagnostic tool, successfully 
signaling that an explicit treatment of SOC response is strictly required for such 
systems. For the other, less demanding systems, we see excellent agreement with the 
data reported by \citet{pennanen_density_2005} and \citet{franzke_paramagnetic_2024}, 
confirming the validity of the model within its intended boundaries.

\begin{table*}
	\centering
		\caption{Calculated isotropic nuclear magnetic shieldings $\sigma_{\text{iso}}^I = \frac{1}{3}(\sigma_{xx}^I + \sigma_{yy}^I + \sigma_{zz}^I)$ 
and chemical shifts $\delta^I$ for the $^{1}\text{H}$ and $^{13}\text{C}$ nuclei in selected metallocenes. 
The spin contribution is represented by $\sigma_{\text{iso}}^{\mathbf{S}(\text{ZS})}$, $\sigma_{\text{iso}}^{\mathbf{S}(\text{Z})}$, 
and $\sigma_{\text{iso}}^{\mathbf{S}(\text{NR})}$, evaluated using the ZORA spin current density, 
Eq.~\eqref{ZORAtttcurrent}, with the approximate treatment of SOC (ZS), without SOC (Z) 
and in the non-relativistic limit (NR), Eq. (\ref{totalcurrent}) (without last term), respectively. 
The orbital contribution $\sigma_{\text{iso}}^{\mathbf{B}}$ 
is evaluated at the non-relativistic level using the CSGT approach. Chemical shifts, $\delta^{I} = \sigma_{\text{ref}}^I - \sigma^I$, are computed 
with respect to ferrocene, where $\sigma_{\text{ref}}^{\text{H}} = 27.54$~ppm 
and $\sigma_{\text{ref}}^{\text{C}} = 101.34$~ppm. All values are reported in ppm.}
	\vspace{3mm}
\begin{tabular}{|l|c|c|c|c|c|c|c|c|c|c|c|c|}
        \hline
        Molecule & $I$ & $\sigma_{\text{iso}}^{\mathbf{B}}$  & $\sigma_{\text{iso}}^{\mathbf{S}(\text{ZS})}$  & $\sigma_{\text{iso}}^{\mathbf{S}(\text{Z})}$  & $\sigma_{\text{iso}}^{\mathbf{S}(\textrm{NR})}$  & $\sigma_{\text{iso}}^{\mathbf{S}(\text{ZS})+\mathbf{B}}$& $\sigma_{\text{iso}}^{\mathbf{S}(\text{Z})+\mathbf{B}}$&$\sigma_{\text{iso}}^{\mathbf{S}(\text{NR})+\mathbf{B}}$&$\delta^{\mathbf{S}(\text{ZS})+\mathbf{B}}$&$\delta^{\mathbf{S}(\text{Z})+\mathbf{B}}$&$\delta^{\mathbf{S}(\text{NR})+\mathbf{B}}$ &$\delta_{\textrm{exp}}$ \\ \hline
        \multirow{2}{*}{\ch{V[C5H5]2}}  & $^1$H & 25.99 & $-$382.12 & $-$382.35 & $-$377.94 & $-$356.13 & $-$356.36 & $-$351.95 & 383.67 & 383.90 & 379.49 & 314.58\\ \cline{2-13}
         & $^{13}$C & 74.32 & 388.13 & 388.05 & 373.81 & 462.45 & 462.36 & 448.13 & $-$361.11 & $-$361.02 & $-$346.79 & -\\ \hline
        \multirow{2}{*}{\ch{Cr[C5H5]2}} & $^1$H & 25.89 & $-$337.84 & $-$338.18 & $-$334.70 & $-$311.95 & $-$312.29 & $-$308.81 & 339.50 & 339.83 & 336.36 & 316.52\\ \cline{2-13}
         & $^{13}$C & 65.53 & 307.73 & 307.60 & 292.13 & 373.26 & 373.13 & 357.66 & $-$271.92 & $-$271.79 & $-$256.32 & -\\ \hline
        \multirow{2}{*}{\ch{Mn[C5H5]2}} & $^1$H & 25.03 & 13.96 & 13.99 & 21.35 & 38.99 & 39.02 & 46.37 & $-$11.44 & $-$11.47 & $-$18.83 & 23.3\\ \cline{2-13}
         & $^{13}$C & 61.05 & $-$1656.76 & $-$1658.54 & $-$1652.49 & $-$1595.71 & $-$1597.49 & $-$1591.44 & 1697.05 & 1698.83 & 1692.78 & 1274\\ \hline
        \multirow{2}{*}{\ch{Co[C5H5]2}} & $^1$H & 27.13 & 55.08 & 55.47 & 55.38 & 82.21 & 82.60 & 82.50 & $-$54.67 & $-$55.05 & $-$54.96 & $-$54.8\\ \cline{2-13}
         & $^{13}$C & 89.67 & $-$697.60 & $-$702.84 & $-$698.10 & $-$607.93 & $-$613.17 & $-$608.42 & 709.27 & 714.51 & 709.76 & -\\ \hline
        \multirow{2}{*}{\ch{Ni[C5H5]2}} & $^1$H & 26.13 & 253.63 & 253.95 & 258.65 & 279.76 & 280.08 & 284.78 & $-$252.21 & $-$252.54 & $-$257.23 & $-$257.44\\ \cline{2-13}
         & $^{13}$C & 82.63 & $-$1699.54 & $-$1703.32 & $-$1679.57 & $-$1616.92 & $-$1620.70 & $-$1596.94 & 1718.26 & 1722.04 & 1698.28 & -\\ \hline
        \multirow{2}{*}{\ch{Rh[C5H5]2}} & $^1$H & 26.58 & 94.99 & 99.58 & 101.55 & 121.57 & 126.16 & 128.12 & $-$94.02 & $-$98.62 & $-$100.58 & -\\ \cline{2-13}
         & $^{13}$C & 86.13 & $-$721.87 & $-$761.05 & $-$743.84 & $-$635.74 & $-$674.93 & $-$657.71 & 737.08 & 776.27 & 759.05 & -\\ \hline
    \end{tabular}
	\label{metalloceni}
\end{table*}

\begin{table*}[htbp!]
	\centering
	\caption{Calculated isotropic nuclear magnetic shieldings $\sigma_{\text{iso}}^I = \frac{1}{3}(\sigma_{xx}^I + \sigma_{yy}^I + \sigma_{zz}^I)$ 
and chemical shifts $\delta^I$ for the $^{1}\text{H}$ and $^{13}\text{C}$ nuclei in N6 nitroxide radical (\ch{C8H15N2O2}). 
The spin contribution is represented by $\sigma_{\text{iso}}^{\mathbf{S}(\text{ZS})}$, $\sigma_{\text{iso}}^{\mathbf{S}(\text{Z})}$, 
and $\sigma_{\text{iso}}^{\mathbf{S}(\text{NR})}$, evaluated using the ZORA spin current density, 
Eq.~\eqref{ZORAtttcurrent}, with the approximate treatment of SOC (ZS), without SOC (Z) 
and in the non-relativistic limit (NR), Eq. (\ref{totalcurrent}) (without last term), respectively.   
The orbital contribution $\sigma_{\text{iso}}^{\mathbf{B}}$ is evaluated at the non-relativistic level using the CSGT approach. Chemical shifts, $\delta^{I} = \sigma_{\text{ref}}^I - \sigma^I$, are computed with respect to TMS, 
where $\sigma_{\text{ref}}^{\text{H}} = 31.38$~ppm and $\sigma_{\text{ref}}^{\text{C}} = 178.98$~ppm. 
All values are reported in ppm. The same atom labels as in Ref.~\citenum{pennanen_density_2005} have been, see Figure \ref{N6Fig}}
	\vspace{3mm}
\begin{tabular}{|l|c|c|c|c|c|c|c|c|c|c|c|}
        \hline
        & $\sigma_{\text{iso}}^{\mathbf{B}}$  & $\sigma_{\text{iso}}^{\mathbf{S}(\text{ZS})}$  & $\sigma_{\text{iso}}^{\mathbf{S}(\text{Z})}$  & $\sigma_{\text{iso}}^{\mathbf{S}(\textrm{NR})}$  & $\sigma_{\text{iso}}^{\mathbf{S}(\text{ZS})+\mathbf{B}}$& $\sigma_{\text{iso}}^{\mathbf{S}(\text{Z})+\mathbf{B}}$&$\sigma_{\text{iso}}^{\mathbf{S}(\text{NR})+\mathbf{B}}$&$\delta^{\mathbf{S}(\text{ZS})+\mathbf{B}}$&$\delta^{\mathbf{S}(\text{Z})+\mathbf{B}}$&$\delta^{\mathbf{S}(\text{NR})+\mathbf{B}}$ &$\delta_{\textrm{exp}}$  \\
        \hline
         C$_1^{\prime}$ & 171.42 & $-$643.13 & $-$643.08 & $-$640.66 & $-$471.71 & $-$471.66 & $-$469.24 & 650.69 & 650.64 & 648.22 & 466.0 \\ \hline
        C$_2$ & 23.27 & 4078.29 & 4074.31 & 4053.42 & 4101.56 & 4097.58 & 4076.69 & $-$3922.57 & $-$3918.60 & $-$3897.71 & / \\ \hline
        C$_4$ & 100.17 & 700.01 & 704.75 & 702.55 & 800.18 & 804.92 & 802.73 & $-$621.20 & $-$625.94 & $-$623.75 & $-$635.0 \\ \hline
        C$_5$ & 99.08 & 718.47 & 714.71 & 712.51 & 817.55 & 813.79 & 811.59 & $-$638.57 & $-$634.81 & $-$632.61 & $-$670.0 \\ \hline
        C$\alpha_{4\textrm{ax}}$ & 152.68 & $-$1047.51 & $-$1047.91 & $-$1042.84 & $-$894.83 & $-$895.23 & $-$890.16 & 1073.82 & 1074.22 & 1069.15 & 1135.0 \\ \hline
        C$\alpha_{4\textrm{eq}}$& 158.66 & $-$519.16 & $-$519.51 & $-$517.28 & $-$360.50 & $-$360.85 & $-$358.62 & 539.48 & 539.83 & 537.60 & 573.0 \\ \hline
        C$\alpha_{5\textrm{ax}}$ & 152.71 & $-$1032.69 & $-$1032.55 & $-$1027.57 & $-$879.98 & $-$879.83 & $-$874.85 & 1058.96 & 1058.82 & 1053.83 & 1170.0 \\ \hline
        C$\alpha_{5\textrm{eq}}$ & 158.52 & $-$500.73 & $-$501.00 & $-$498.87 & $-$342.21 & $-$342.47 & $-$340.35 & 521.19 & 521.46 & 519.33 & 650.0 \\ \hline
        H$_{4/5\textrm{ax}}$ & 30.25 & 14.52 & 14.53 & 14.59 & 44.77 & 44.78 & 44.84 & $-$13.39 & $-$13.40 & $-$13.46 & $-$13.1 \\ \hline
        H$_{4/5\textrm{eq}}$ & 30.16 & 18.54 & 18.56 & 18.57 & 48.70 & 48.72 & 48.73 & $-$17.32 & $-$17.34 & $-$17.34 & $-$13.9 \\ \hline
        H$_1^{\prime}$ & 29.45 & 358.80 & 358.83 & 358.78 & 388.25 & 388.28 & 388.23 & $-$356.86 & $-$356.80 & $-$356.85 & $-$230.6 \\ \hline
    \end{tabular}
	\label{N6}
\end{table*}

Since there are some differences in computational settings between our calculations and those in the literature (in terms of basis sets and density functional approximations), in addition to the differences in the formalism, we provide below a theoretical comparison to results by~\citet{franzke_paramagnetic_2024}, which are based on the X2C Hamiltonian and include the full response to the external magnetic field. 
The comparison is carried out on five of the seven molecules listed in Table~\ref{shmolecules},
adopting the same geometries and basis sets as in the table.  
We excluded \ch{C8H15N2O2} and \ch{Rh[C5H5]2}, since they were not originally part of the dataset from  Ref.~[\citenum{franzke_paramagnetic_2024}].
For this comparison, we adopted the same spherical basis sets as used by \citet{franzke_paramagnetic_2024}, which we then converted into Cartesian Gaussian-type orbitals (GTOs) using the default Gaussian 16 transformation routine. 
Despite this conversion, the total energies and the resulting electronic properties remain numerically identical to those obtained using a spherical basis set, ensuring that no bias is introduced and allowing for a rigorous validation against the reference data. 
Finally, we have restricted ourselves to the B3LYP functional.

Magnetic properties were evaluated using Gaussian 16. Specifically, the orbital contributions were obtained employing the CSGT method, as mentioned earlier.
To account for scalar relativistic effects and retrieve the spin-related contributions, the DKH Hamiltonian was invoked via the \texttt{int=dkh} keyword. The analysis was conducted by processing electronic structure data exported into \texttt{.wfx} files using the SYSMOIC software package, utilizing the newly introduced module \cite{monaco_program_2021}. 
For each molecular system, two distinct \texttt{.wfx} files were generated to separately treat the orbital and spin contributions.  Note that pseudo-contact shifts were not considered in the analysis presented in this section. 

\begin{figure}[htbp!]
    \centering
    \includegraphics[width=1.0\linewidth]{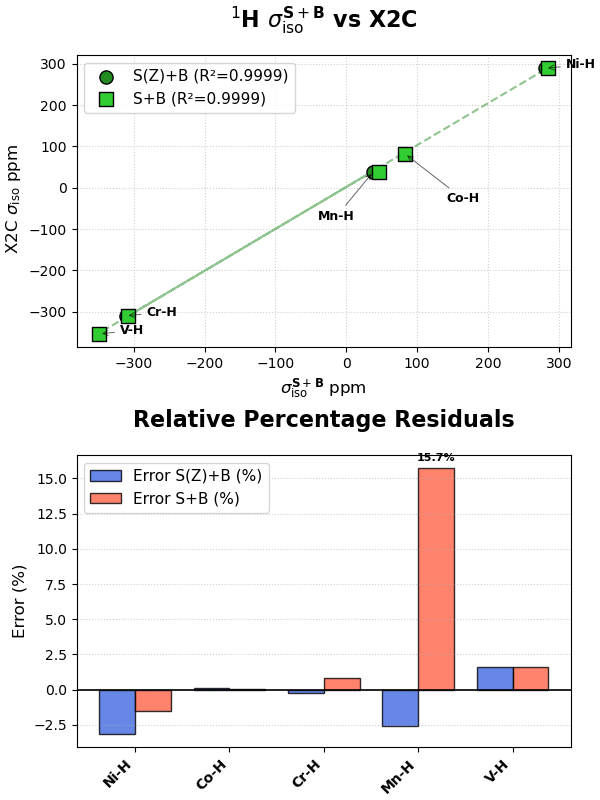}
    \caption{Comparison of calculated $^1$H isotropic shielding constants ($\sigma_{\text{iso}}$) against the X2C reference data from \citet{franzke_paramagnetic_2024}. 
    Top: Linear correlation plot for the non-relativistic (S+B, green squares) and scalar relativistic (S(Z)+B, green circles) approaches. 
    Bottom: Relative percentage residuals for both methods.}
    \label{fig:h1_comparison}
\end{figure}

The comparison of $^1$H isotropic shielding constants ($\sigma_{\text{iso}}$) against the X2C results is summarized in 
Figure~\ref{fig:h1_comparison}. Our results demonstrate that the inclusion of scalar relativistic effects is crucial for an accurate description of these systems, accounting for the scalar HALA effect. As shown in the residuals plot, the non-relativistic approach (S+B, in red) exhibits significant deviations, with a maximum error of approximately 16\% for the Mn-H complex due to spin contributions. 

In contrast, the scalar relativistic treatment (S(Z)+B, in blue) drastically reduces these discrepancies. The residuals for the S(Z)+B method remain consistently low across the entire set ($R^2 = 0.9999$), indicating an almost ideal agreement with the reference X2C data. The small residual differences in the S(Z)+B implementation can be attributed to the different treatment of the relativistic Hamiltonian (DKH/ZORA vs. X2C) and intrinsic differences in the property kernels. 

For the Mn system, the relative errors for both non-relativistic and scalar relativistic Hamiltonians are significantly larger than those for the other systems.
\citet{franzke_paramagnetic_2024} similarly found that the calculated
$^{1}$H pNMR shifts of the same compound did not follow the experimental trend as the complexes of V, Cr and Co. The authors attributed this difference in behavior to deficiencies of DFT to properly describe the hyperfine coupling constant tensors for this system. Given the very large differences in residuals already observed for a non-relativistic Hamiltonian, and the fact that electron correlation and relativistic effects are known to be non-additive, 
we speculate that in our case these differences are being mainly driven by the deficiencies of DFT. Unfortunately we are currently not in a position to  investigate this issue further, as such an analysis would likely require a wavefunction-based treatment, which falls outside the scope of the present work.

\begin{figure}[htbp!]
    \centering \includegraphics[width=1.0\linewidth]{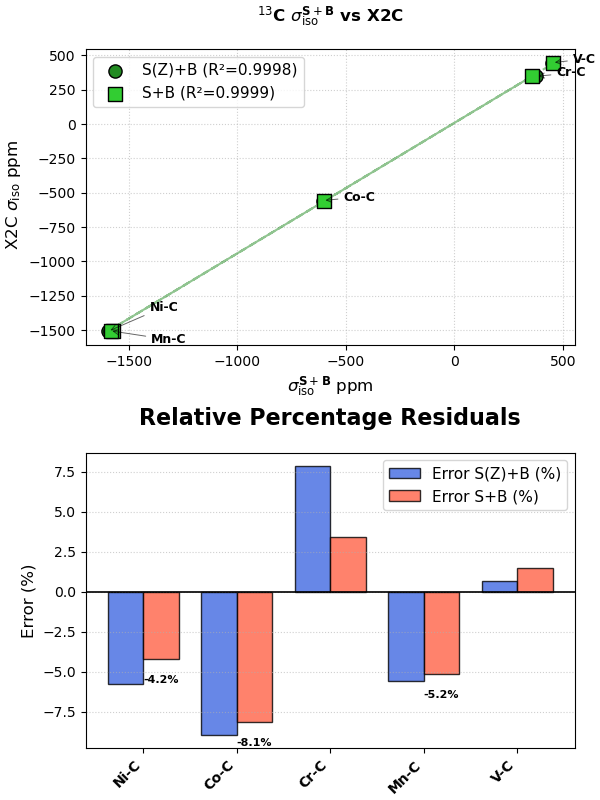}
    \caption{Comparison of calculated \textsuperscript{13}C isotropic shielding constants ($\sigma_{\text{iso}}$) against X2C reference data. 
    Top: Linear correlation plot. 
    Bottom: Relative percentage residuals.}
\label{fig:c13_comparison}
\end{figure}

The comparison for $^{13}$C isotropic shielding constants is illustrated in Figure~\ref{fig:c13_comparison}. 
In this case, the transition from the non-relativistic to the scalar relativistic framework does not yield a systematic improvement across all systems. While the correlation remains high ($R^2 \approx 0.999$), the relative residuals for both methods are comparable (5--10\%). For systems such as Ni-C and Co-C, the scalar relativistic correction does not significantly reduce the discrepancy. 
This suggests that for carbon nuclei in these environments, the dominant source of deviation might reside in higher-order relativistic contributions (e.g., SOC) or subtle differences in the property kernels. Nevertheless, the overall agreement remains within a physically acceptable range for paramagnetic NMR benchmarking. 

\begin{figure}[htbp!]
    \centering \includegraphics[width=0.9\linewidth]{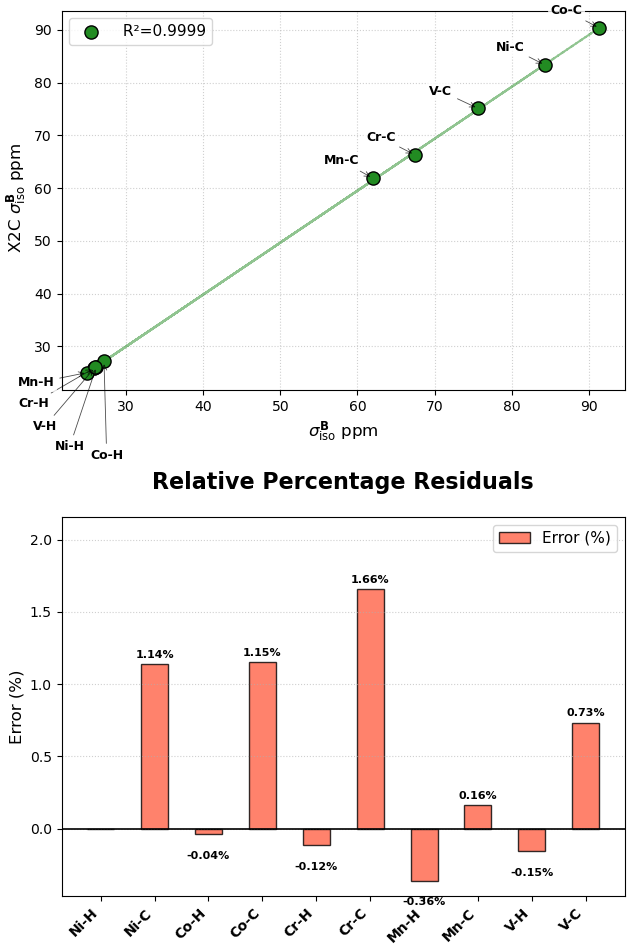}
    \caption{Correlation and absolute residuals between the orbital contributions ($\sigma_{\text{iso}}^{\text{B}}$) calculated with the CSGT method and the X2C reference data.} \label{fig:orbital_comparison}
\end{figure}

Finally, we compared the orbital contributions ($\sigma_{\text{iso}}^{\text{B}}$) obtained via CSGT with the corresponding X2C values (Figure \ref{fig:orbital_comparison}).
 The two methods yield nearly identical results ($R^2 = 0.9999$, slope $= 0.9839$). The absolute residuals remain within a narrow range of $\pm 0.5$~ppm, demonstrating that the orbital component is consistently captured by the CSGT approach, matching the more demanding X2C treatment and justifying this partitioned scheme as a reliable approximation. 

\clearpage

\subsection{Nuclear Hyperfine Coupling Constants}

The isotropic components of the nuclear hyperfine coupling constants (NHCCs) $A_{\text{iso}}$ with the current density approach have been obtained for all atoms in the 3d and 4d metallocenes systems described above. Our results (in MHz) are reported in Table \ref{tab:hfc_results}. For $^{1}$H and $^{13}$C, the reported values are averaged over all equivalent nuclei in the system.
\begin{table}[htbp!]
\centering
\caption{Isotropic nuclear hyperfine coupling constants $A_{\text{iso}}$ (MHz) for various metallocenes, computed at non-relativistic S, scalar relativistic S(Z), and spin-orbit coupling relativistic S(ZS) levels of theory using equation (\ref{hyperfineA}). Analytical results are provided for the non-relativistic case S(A). Values in parenthesis denote scalar relativistic ($\Delta_\text{SR} = \text{S(Z)}-\text{S}$, displayed in column S(Z)) and spin-orbit ($\Delta_\text{SO} = \text{S(SZ)}-\text{S(Z)}$, displayed in column S(ZS)) effects.}
	\label{tab:hfc_results}
	\begin{tabular}{lccrrr}
		\toprule
		Molecule & I & S\,\,\, & S(A) & S(Z) & S(ZS)\\
		\midrule
        \ch{V[C5H5]2}& $^{51}$V & $-$25.606 &$-$25.607 &$-$13.995 & $-$14.042 \\
        &            &            &           & (11.611) & ($-$0.047) \\
        & $^{13}$C & $-$0.710 &$-$0.709 & $-$0.737 & $-$0.737 \\
        &            &            &           & ($-$0.027) & (0.000) \\
        	& \,\,$^1$H&+2.852 &+2.852 & +2.886 & +2.885 \\
        &            &            &           & (0.034) & ($-$0.001) \\
        \addlinespace
        \ch{Cr[C5H5]2}
        & $^{53}$Cr &+17.786 &+17.786 &+14.508 &+14.487 \\
        &            &            &           & ($-$3.278) & ($-$0.021) \\
        & $^{13}$C & $-$1.040 &$-$1.040 &$-$1.095 & $-$1.096 \\
        &            &            &           & ($-$0.055) & ($-$0.001) \\
        & \,\,$^1$H&+4.736  &+4.736 &+4.785 & +4.784 \\
        &            &            &           & (0.049) & ($-$0.001) \\
    \addlinespace
        \ch{Mn[C5H5]2}& $^{55}$Mn & $-$141.392 &$-$141.389 & $-$127.464 & $-$125.749 \\
        &            &            &           & (13.928) & (1.715) \\
        & $^{13}$C & +1.759 &+1.759 & +1.766  & +1.765 \\
        &            &            &           & (0.007) & ($-$0.001) \\
        	& \,\,$^1$H& $-$0.090 &$-$0.090 & $-$0.059 & $-$0.059 \\
        &            &            &           & (0.031) & (0.000) \\
		\addlinespace
		 \ch{Co[C5H5]2}
        & $^{59}$Co  & $-$282.781 &$-$282.780 & $-$263.183 & $-$264.901 \\
        &            &            &           & (19.598) & ($-$1.718) \\
		& $^{13}$C   &     +6.625 &+6.625  &   +6.670 &     +6.648 \\
        &            &            &           & (0.045) & ($-$0.022) \\
		& \,\,$^1$H  &   $-$2.090 &$-$2.090  & $-$2.093 &   $-$2.087 \\
        &            &            &           & ($-$0.003) & (0.006) \\
		\addlinespace
        \ch{Ni[C5H5]2}
        & $^{61}$Ni  &   +105.106 & +105.112  & +98.304 &   +98.462 \\
        &            &            &           & ($-$6.802) & (0.158) \\
        & $^{13}$C   &     +5.977 & +5.977  &  +6.062 &    +6.066 \\
        &            &            &           & (0.085) & (0.004) \\
        & \,\,$^1$H  &   $-$3.660 & $-$3.660  & $-$3.594 &  $-$3.599 \\
        &            &            &           & (0.066) & ($-$0.005) \\
 		\addlinespace
		 \ch{Rh[C5H5]2}
        & $^{103}$Rh &+82.698     &+82.696   &+106.789    &+106.050    \\
        &            &            &           & (24.091) & ($-$0.739) \\
		& $^{13}$C   &+7.059     &+7.059    &+7.223    &+7.119    \\
        &            &            &           & (0.164) & ($-$0.104) \\
		& \,\,$^1$H  &$-$3.832   &$-$3.832  &$-$3.758  &$-$3.725  \\
        &            &            &           & (0.074) & (0.033) \\		\bottomrule
	\end{tabular}
\end{table}

The magnitude of relativistic effects on $A_{\text{iso}}$ is rather small for $^{1}$H, typically $<$ 0.05 MHz for the 3d complexes but reaching about 0.1 MHz for Rhodocene, and essentially due to scalar relativistic effects. 

The overall effect of relativity for $^{13}$C turns out to be similar in magnitude to that of $^{1}$H. While for the 3d metals this is due to the fact that SOC contributions ($\Delta_\text{SO}$) are very small (typically $<$ 0.01 MHz), and when they are more important, they are of similar magnitude and opposite sign to the scalar relativistic ($\Delta_\text{SR}$) contributions, as one can see from the values of the $\Delta_\text{SR}$ and $\Delta_\text{SO}$ presented in Table \ref{tab:hfc_results} (see table caption for definitions). This explains why for Rhodocene the $^{13}$C NHCC is of similar magnitude to that for $^{1}$H.

For the metal center NHCCs, we see a similar trend in that scalar and SOC contributions typically have opposite signs, though in this case contributions for scalar relativistic effects are typically much larger than those from SOC and result in sizable contributions. With respect to the difference in signs, the \ch{Mn[C5H5]2} system is the exception, perhaps due to the difficulties in accounting for electron correlation with approximate density functional approximations as discussed above for shieldings, with the same sign for scalar and SOC effects making it the 3d metal system with the largest NHCC. At the same time, for 
\ch{Ni[C5H5]2} and \ch{Cr[C5H5]2} scalar relativistic effects are smaller in magnitude and with a negative sign, so that the NHCCs are actually reduced compared to the non-relativistic result. Finally, as expected the heaviest system, \ch{Rh[C5H5]2}, shows the largest change in NHCCs. 

Taken together, these results make it clear that while it is indispensable to take into account both scalar relativistic and SOC effects for NHCCs, for 3d and even 4d elements a scalar relativistic treatment captures most of the effects. 

\begin{figure}[htbp!]
    \centering
\includegraphics[width=1.0\linewidth]{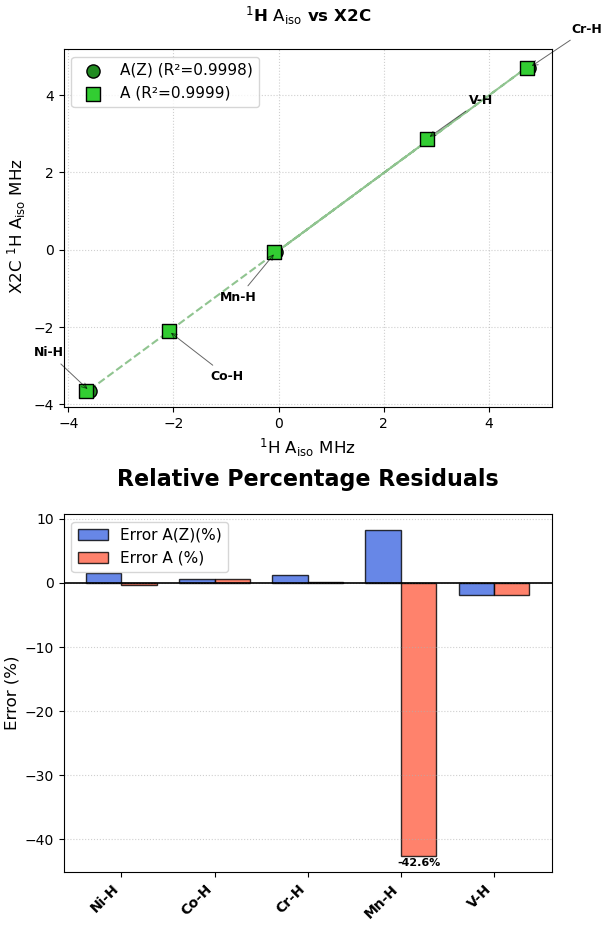}
    \caption{Comparison of calculated \textsuperscript{1}H isotropic hyperfine coupling tensor against X2C reference data. Top: Linear correlation plot. 
    Bottom: Relative percentage residuals.}
    \label{fig:h1_a_comparison}
\end{figure}

\begin{figure}[htbp]
    \centering
\includegraphics[width=1.0\linewidth]{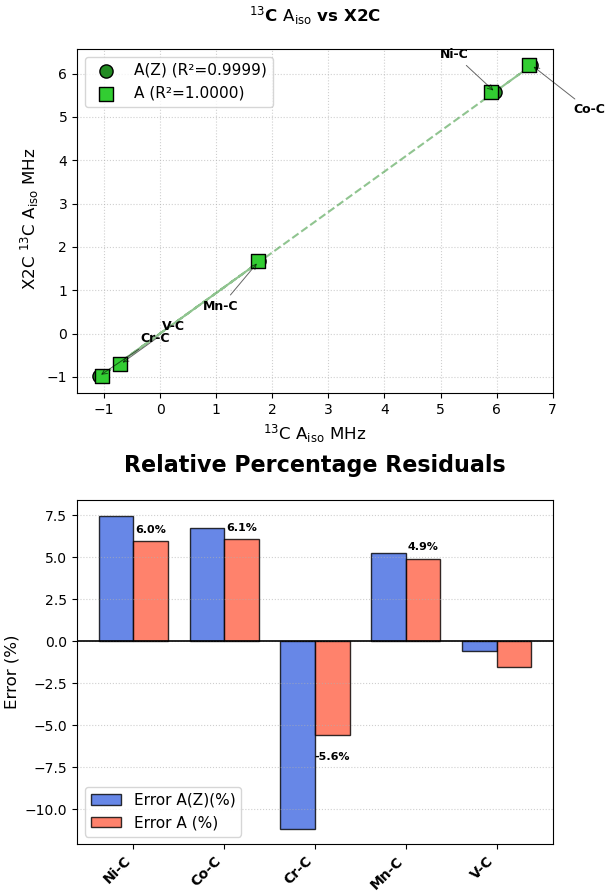}
    \caption{Comparison of calculated \textsuperscript{13}C isotropic hyperfine coupling tensor against X2C reference data. Top: Linear correlation plot. 
    Bottom: Relative percentage residuals.}
    \label{fig:c13_a_comparison}
\end{figure}

As in the case of nuclear magnetic shieldings, for $^1$H and $^{13}$C  nuclear hyperfine isotropic tensors for the molecules under consideration, we observe from Figures \ref{fig:h1_a_comparison} and \ref{fig:c13_a_comparison} that our result compare favorably to those obtained with the X2C approach by \citet{franzke_paramagnetic_2024}, indicating that both approach capture the same dominant physical information. This is expected being the hyperfine coupling tensor directly related to the spin contribution of nuclear magnetic shielding tensor.

\clearpage

\subsection{Magnetizabilities}

In this section we present our results for the isotropic components of the magnetizability tensors, focusing on three molecules containing light elements only (\ch{O2}, \ch{NO} and \ch{NO2}) for which experimental results in the gas phase are available. The results, presented in 
Table~\ref{MAG1}.

Prior the calculation of the magnetizabilties, we carried out geometry optimizations and frequency calculations at the B3LYP/X2C-QZVPall-s level of theory using Gaussian 16~\cite{g16}. 
Vibrational analysis confirmed that all optimized structures correspond to true local minima, as indicated by the absence of imaginary frequencies. On top of that, we carried out calculations of NMR properties with the CSGT method.

From the results on Table~\ref{MAG1}, our approach shows a very good agreement with experiment. We observe that the spin contribution completely dominates the orbital contribution for these systems. Although a set of three molecules is not sufficient for a comprehensive statistical analysis, this suggests the model presented here is highly promising, showing excellent agreement with Curie's law regarding the paramagnetic contribution to the magnetizability. 

\begin{table*}[!ht]
	\centering
	\caption{Calculated and experimental magnetizability components for \ch{O2}, \ch{NO} and \ch{NO2} at the B3LYP/X2C-QZVPall-s level of theory. Values represent the isotropic average of the tensor, defined as $\chi = \frac{1}{3}(\chi_{xx} + \chi_{yy} + \chi_{zz})$. The spin and orbital contributions are denoted by $\chi^{\mathbf{S}}$ and $\chi^{\mathbf{B}}$ respectively. Temperature effects are accounted for through Eq.~\ref{spinmagn}. Units are ppm cm$^3$ mol$^{-1}$.}
	\vspace{3mm}
	\begin{tabular}{lcccccccc} 
		\toprule
		\textbf{Molecule} & \textbf{2S+1} & \textbf{T (K)} &  $\chi^{\mathbf{S}}$ & $\chi^{\mathbf{B}}$ & $\chi^{\mathbf{S}+\mathbf{B}}$ & \textbf{Exp. } \\
        \midrule
		\ch{O2}  & 3 & 295.75 &+3390.3  & $-$11.0 & +3379.3 & +3330\cite{Havens1932} \\
		\ch{NO}  & 2 & 298.15 &+1261.1  & +65.8 & +1326.9 & +1461\cite{CRC_Handbook_2024} \\
        \ch{NO2}  & 2 & 295.75 &+1271.3  &$-$12.0  &+1259.3  &+1249\cite{Havens1932} \\
		\bottomrule
	\end{tabular}
	\label{MAG1}
\end{table*}
\subsection{Visualization of Spin current Densities and Property Densities in Real Space}
An additional attractive feature of expressing molecular magnetic properties in terms of current densities \cite{soncini_charge_2007,zanasi_ring_1995}, together with their decomposition into spin-independent and spin-dependent components, is the ability to visualize the resulting quantities in real space. Such visualizations allow for a topological analysis of the associated vector fields, thereby providing further characterization of the underlying physical properties of the systems.
One should, however, keep in mind some important differences for the topological analysis of the induced currents between closed and open-shell systems. For closed shell systems, the topological analysis is directly related to the orbital contributions to the magnetic response of the systems, with no contribution from the spin unless spin-orbit coupling effects are sufficiently strong---that is, for systems at the bottom of the periodic table such as lanthanides and actinides, 5d transition metals, etc. 
For open-shell systems, on the other hand, spin contributions will always be important, and the macroscopic spin contribution to NMR parameters is intrinsically temperature-dependent, since it results from a statistical (Boltzmann) average over the states within the spin multiplet. 
This means that, in order to arrive at a qualitatively accurate description of the physical system, one has to compose a vector field resulting from contributions from all populated spin states at a given temperature before proceeding to any topological analysis, otherwise one risks analyzing an artifact.

\begin{figure}[!ht]
	\centering
\includegraphics[width=0.9\linewidth]{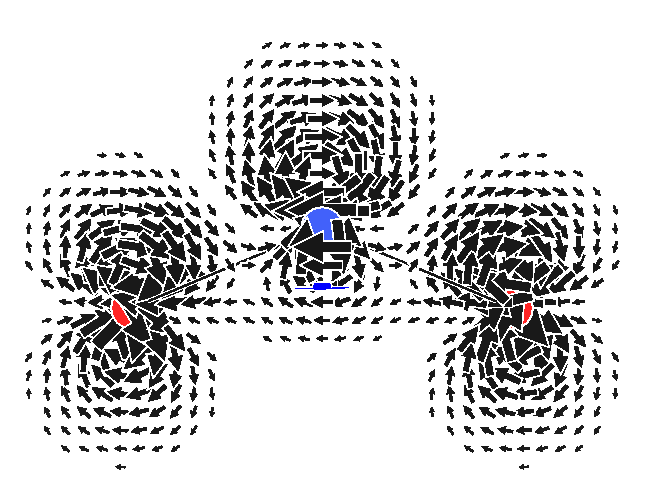}
	\caption{ZORA Spin Current density map of \ch{NO2} ($2S+1=2$) induced by a magnetic field $B_x\mathbf{e}_x$ pointing toward the reader for the maximum polarized state $N_\alpha>N_\beta$.}
	\label{SCUR2}
\end{figure}

To illustrate this point, consider the \ch{NO2} ($2S+1=2$) molecule, which has two spin states, a maximally polarized state with $M_S = S$ ($N_\beta < N_\alpha$) whose magnetically induced current density map (induced by a magnetic field $B_x\mathbf{e}_x$) is shown in Figure~\ref{SCUR2}, and
a minimally polarized state with $M_S = -S$ ($N_\beta > N_\alpha$) that would show a reversed spin current pattern, though not equal to that of the maximally polarized state, thus resulting in the net contribution found on Table~\ref{MAG1}) 
(we note that for technical reasons we cannot at this time carry out DFT calculations for the minimally polarized state, and with that present the corresponding spin current density map).

\begin{figure}[!ht]
	\centering
	\includegraphics[width=0.5\linewidth]{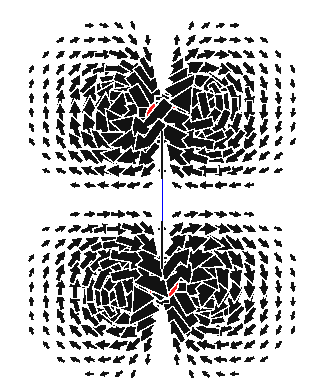}
	\caption{ZORA Spin Current density maps induced by a magnetic field $B_x\mathbf{e}_x$ pointing toward the reader, for the maximum polarized state $N_\alpha>N_\beta$, in \ch{O2} ($2S+1=3$).}
	\label{SCUR}
\end{figure}

While in the case above it may be that the analysis of the maximally polarized state could still not be too far off from that of the ensemble, the current density maps for systems with more spin states, such as the oxygen molecule shown in Figure~\ref{SCUR} (for the maximally polarized state) may be far removed from the behavior of the physical system.

\begin{figure*}[hbpt!]
	\centering
\includegraphics[width=1.0\textwidth]{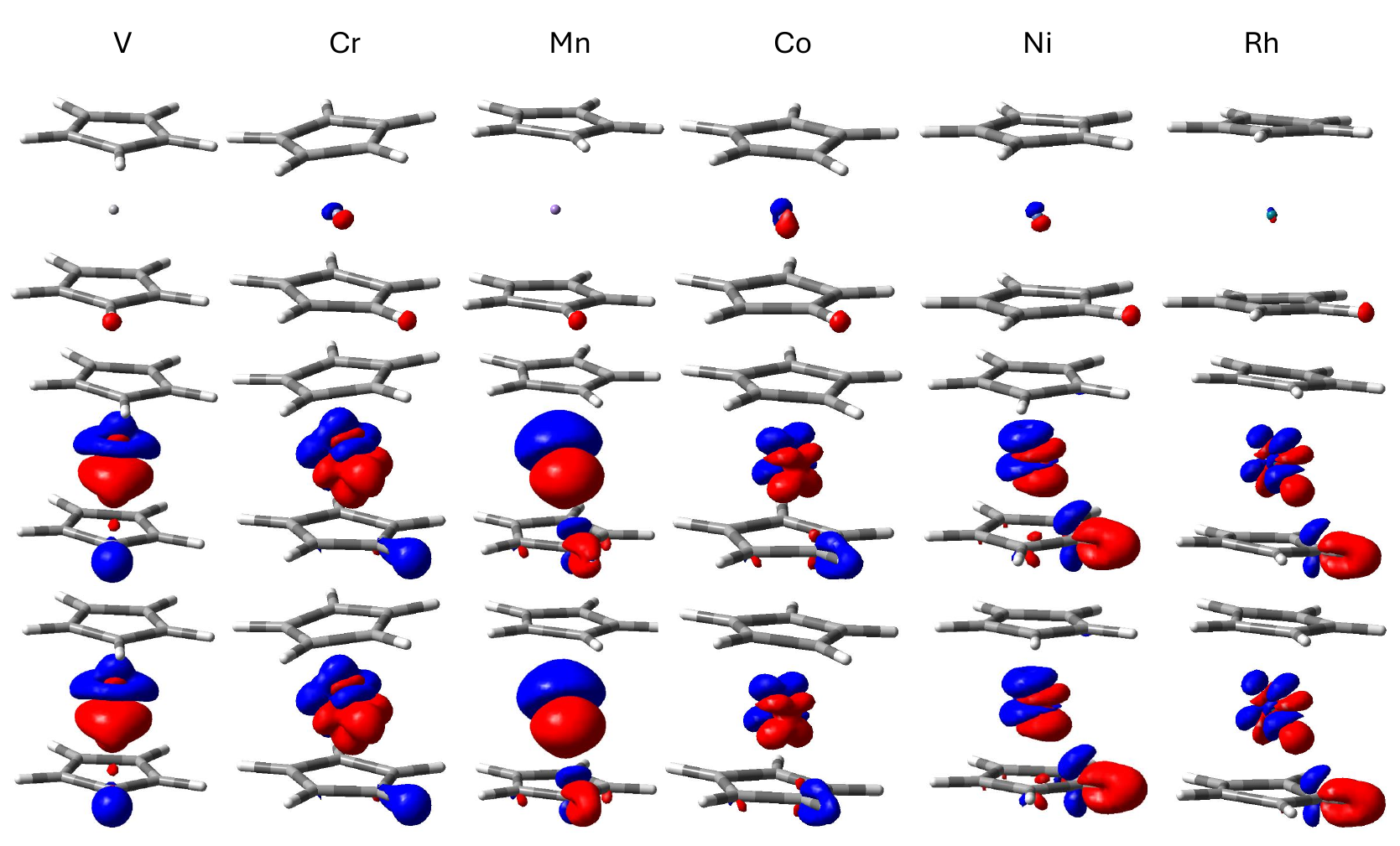}
\caption{Isotropic contribution to the $^1$H nuclear magnetic shielding density for metallocenes at 298.15 K. Top: orbital shielding densities, defined from 
Eq.~\ref{orbshield} using the CTOCD method (with the CSGT option of the shift function); middle: the spin contribution, defined from 
Eq.~\eqref{spinshield} using a scalar relativistic spin density throughout the first term of the ZORA spin current; and bottom: the total shielding density (sum of the two contributions). Isosurface values are set at $\pm 10$~ppm/$a_0^3$. Note that different protons have different distributions due to anisotropic character of the spin density on the metal center.}
\label{shielddens}
\end{figure*}

An alternative to the analysis of current density maps is provided by property densities, whose definition, as follows from Eqs.~\eqref{spinshield} for shieldings and Eq.~\eqref{spinmagn} for magnetizabilities, 
already involve an average over contributions from different thermally populated 
spin states. 
As illustration, we present in Figure~\ref{shielddens} the nuclear magnetic shielding densities for one of the $^1$H of metallocenes. We note that, in order to obtain the numerically integrated properties in Table \ref{metalloceni}, it would be necessary to sum over the contributions of all hydrogen atoms. 
With this, is it not strictly possible to connect these plots to experiment.
From the figure, is it clear that spin contributions are consistently dominant, accounting for nearly the entire total nuclear magnetic shielding in all investigated systems. 
While this trend is consistent with that reported in Table  \ref{metalloceni},  the present analysis allows us to observe changes in the spatial extent of the contributions around the hydrogen atom. In particular, the contribution on the bonded carbon atom increases as the metal center becomes heavier.

\section{Conclusions}
\label{Conclusions}

In this work, we presented a novel computational approach for the evaluation of pNMR chemical shifts, magnetizabilities and nuclear hyperfine coupling costant tensors in paramagnetic systems. 

A key advantage of the proposed method is that it circumvents the explicit calculation of the g-tensor and the zero-field splitting (ZFS) Hamiltonian~\cite{van_den_heuvel_nmr_2013,autschbach_nmr_2015,martin_temperature_2015,martin_erratum_2016,franzke_paramagnetic_2024}. This significantly reduces both the computational complexity and the potential sources of error associated with these parameters in traditional frameworks.
This simplification is justified, for the nuclei considered in the molecules analyzed here, because spin–Zeeman and spin–orbit coupling interactions can be neglected to a first approximation, even though they are used in the derivation of the current density expression.

The results demonstrate that the approach provides shielding values in good agreement with those obtained using the \textit{exact} two-component (X2C) method, which currently represents the gold standard for two-component relativistic nuclear magnetic shielding calculations. The robustness of the method is further confirmed by the agreement found with experimental data across the diverse set of transition metal complexes investigated.  

Our novel formulation not only streamlines the prediction of paramagnetic NMR parameters but also establishes a reliable bridge between simplified density-based models and fully relativistic Hamiltonians. The consistency with X2C benchmarks suggests that the present approach accurately captures the essential physics of the spin-dependent interactions, making it a powerful tool for the characterization of complex open-shell molecular systems containing up to 3d transition metals.  

The implementation and detailed analysis of scalar relativistic and spin–orbit coupling effects on the orbital contributions to the shielding and magnetizability tensors will be the subject of a future investigation. All terms up to first order in the applied magnetic field will be included, using the CTOCD technique to ensure gauge-origin independence.

With proper inclusion of spin–orbit coupling and spin–Zeeman interactions, we expect that the proposed method can be straightforwardly extended and effectively applied across the entire periodic table. Such an extension would enable a consistent and accurate description of paramagnetic NMR parameters even for complexes involving the heaviest elements, where scalar and spin–orbit relativistic effects become equally important.

\section{Acknowledgements}

We would like to thank you Profs. R. Zanasi, G. Monaco and P. Lazzeretti for helpful discussions. Financial support from MUR (FARB 2022 and FARB 2023) is gratefully acknowledged. We also thank Ph.D. Lufeng Zou for his helpful support with Gaussian 16.

ASPG acknowledges funding from projects CPER WaveTech, Labex CaPPA (Grant No. ANR-11-LABX-0005-01), ANR SCREECHES (Grant Nos. ANR-24-CE29-0904) the I-SITE ULNE project OVERSEE and MESONM International Associated Laboratory (LAI) (Grant No. ANR-16-IDEX-0004), as well as support from the French national supercomputing facilities (Grant A0210801859) and the HPC center at the University of Lille.

\clearpage

\appendix
\section{Spin Contributions to Electron Current Density}
\label{Section Spin Contributions to Current Density}
In this section, we apply the Landau approach \cite{landau_quantum_2007,summa_molecular_2024} to derive the ZORA contributions to the electron current density originating from the spin-Zeeman and spin-orbit coupling Hamiltonians. 
Let us consider first the one-electron ZORA Hamiltonian for the spin-Zeeman interaction
\begin{equation}
	\hat{\mathrm{h}}(1)=g_{\textrm{e}}\mu_{B}\frac{2m_{\textrm{e}}}{\hbar}\mathcal{K}(\boldsymbol{r})\hat{\boldsymbol{s}} \cdot (\boldsymbol{\nabla}\times\boldsymbol{A})~.
\end{equation}
Replacing $\boldsymbol{A}(\boldsymbol{r})$ with $\boldsymbol{A}(\boldsymbol{r})+\delta \boldsymbol{A}(\boldsymbol{r})$,  we obtain
\begin{equation}
	\small
	\begin{split}
		{H}_{c}+\delta {H}_{c}&=	g_{\textrm{e}}\mu_{B}\frac{2m_{\textrm{e}}}{\hbar}n\int  \left[\Psi^{*}\mathcal{K}\hat{\boldsymbol{s}} \cdot (\boldsymbol{\nabla}\times\boldsymbol{A})\Psi \right]\,d \boldsymbol{X}_{1}\,d \eta_1 \,d^3 r \\
&+g_{\textrm{e}}\mu_{B}\frac{2m_{\textrm{e}}}{\hbar}n\int  \left[\Psi^{*}\mathcal{K}\hat{\boldsymbol{s}} \cdot (\boldsymbol{\nabla}\times\delta\boldsymbol{A})\Psi\right] \,d \boldsymbol{X}_{1}\,d \eta_1 \,d^3 r
	\end{split}
\end{equation}
from which it is possible to rewrite
\begin{equation}
	\delta {H}_{c}=g_{\textrm{e}}\mu_{B}\frac{2m_{\textrm{e}}}{\hbar}n \int (\boldsymbol{\nabla}\times\delta\boldsymbol{A})\cdot  \Psi^{*}\mathcal{K}\hat{\boldsymbol{s}}\,\Psi\, d \boldsymbol{X}_{1}\,d \eta_1 \,d^3 r~,
	\label{ZeemandHc}
\end{equation}
where dependencies  have been omitted for sake of simplicity. 
Now, taking into account the vector identity
\begin{equation}
	\begin{aligned}
		\boldsymbol{\nabla} \times \delta \boldsymbol{A} \cdot \Psi^{*} \mathcal{K}\hat{\boldsymbol{s}} \Psi&= \boldsymbol{\nabla} \cdot(\delta \boldsymbol{A} \times \Psi^{*}\mathcal{K}\hat{\boldsymbol{s}} \Psi)\\&+\delta \boldsymbol{A} \cdot \boldsymbol{\nabla} \times \Psi^{*}\mathcal{K}\hat{\boldsymbol{s}} \Psi
	\end{aligned}
\end{equation}
and applying the divergence theorem for the first term on the r.h.s of the previous equation, considering that the wavefunction goes to zero at infinity, we have
\begin{equation}
	\int  \boldsymbol{\nabla} \cdot(\delta \boldsymbol{A} \times \Psi^{*}\mathcal{K}\hat{\boldsymbol{s}} \Psi) \,d^3 r=0~.
\end{equation}
From this,  
Eq.~\eqref{ZeemandHc} can then be rewritten as
\begin{equation}
	\small
	\delta {H}_{c}=g_{\textrm{e}}\mu_{B}\frac{2m_{\textrm{e}}}{\hbar}n\int \left \{   \boldsymbol{\nabla}\times\left[\int \Psi^{*}\mathcal{K}\hat{\boldsymbol{s}}\Psi\,d \boldsymbol{X}_{1}\,d \eta_1\right]\,  \right \}\cdot\,\delta\boldsymbol{A} \,d^3 r~.
\end{equation}
Considering the vector identity 
\begin{equation}
\boldsymbol{\nabla}\times\left[\mathcal{K}(\boldsymbol{r})\boldsymbol{Q}(\boldsymbol{r})\right]=\mathcal{K}(\boldsymbol{r})\left[\boldsymbol{\nabla}\times\boldsymbol{Q}(\boldsymbol{r})\right]+\left[\boldsymbol{\nabla}\mathcal{K}(\boldsymbol{r})\right] \times \boldsymbol{Q}(\boldsymbol{r})
\end{equation}
and that
\begin{equation}
	\boldsymbol{\nabla}\mathcal{K}(\boldsymbol{r})=\frac{e\mathcal{K}(\boldsymbol{r})\boldsymbol{\nabla}V(\boldsymbol{r})}{2m_{\textrm{e}}c^2-eV(\boldsymbol{r})}=\frac{e}{c^2}\mathcal{K}^2(\boldsymbol{r})\boldsymbol{\nabla}V(\boldsymbol{r})
	\label{Kderiv}
\end{equation}
by using the definition provided in Eq.~\eqref{classic H variation}, we obtain, as contribution to the total induced electron current density vector, the expression 
\begin{equation}
\begin{split}
	\boldsymbol{J}(\boldsymbol{r})=&-g_{\textrm{e}}\mu_{B}\frac{2m_{\textrm{e}}}{\hbar}\mathcal{K}(\boldsymbol{r})\,\boldsymbol{\nabla}\times\boldsymbol{Q}(\boldsymbol{r})\\&-g_{\textrm{e}}\frac{e^2}{c^2}\mathcal{K}^2(\boldsymbol{r})\boldsymbol{\nabla}V(\boldsymbol{r})\times \boldsymbol{Q}(\boldsymbol{r})~.
	\label{magnet-curr}
    \end{split}
\end{equation}
A procedure similar to the ones adopted above to derive 
Eq.~\eqref{magnet-curr} can be applied to the one-electron spin-orbit coupling Hamiltonian 
\begin{equation}
	\begin{split}
	\hat{\mathrm{h}}(1)=&\,g_{\textrm{e}}\frac{e	\mathcal{K}(\boldsymbol{r})}{2m_{\textrm{e}}c^2-eV(\boldsymbol{r})}\sum_{I=1}^{N}\hat{\boldsymbol{s}}\cdot \left [\boldsymbol{\nabla}V^{I}\times \hat{\boldsymbol{\pi}}\right]\\ =&\,g_{\textrm{e}}\frac{e}{c^2}\mathcal{K}^2(\boldsymbol{r})\sum_{I=1}^{N}\hat{\boldsymbol{s}}\cdot \left [\boldsymbol{\nabla}V^{I}\times \hat{\boldsymbol{\pi}}\right]
\end{split}
\end{equation}
Again, substituting $\boldsymbol{A}(\boldsymbol{r})$ with $\boldsymbol{A}+\delta\boldsymbol{A}$, we get
\begin{equation}
	\small
	\begin{split}
		{H}_{c}+\delta {H}_{c}=g_{\textrm{e}}\frac{ne}{c^2}\sum_{I=1}^{N} &\int  \Psi^{*}\mathcal{K}^2\hat{\boldsymbol{s}} \cdot \left[\boldsymbol{\nabla}V^{I}\times\boldsymbol{\hat{\pi}}\right]\Psi \,d \boldsymbol{X}_{1}\,d\eta_1 \,d^3 r \\
		+g_{\textrm{e}}\frac{ne}{c^2}\sum_{I=1}^{N} &\int  \Psi^{*}\mathcal{K}^2\hat{\boldsymbol{s}} \cdot \left[\boldsymbol{\nabla}V^{I}\times e \delta\boldsymbol{A}\right]\Psi\,d\boldsymbol{X}_{1}\,d\eta_1 \,d^3 r~,
	\end{split}
\end{equation}
from which it follows
\begin{equation}
	\delta {H}_{c}=g_{\textrm{e}}\frac{ne}{c^2}\sum_{I=1}^{N} \int  \Psi^{*}\mathcal{K}^2\hat{\boldsymbol{s}} \cdot \left[\boldsymbol{\nabla}V^{I}\times e \delta\boldsymbol{A}\right]\Psi\,d \boldsymbol{X}_{1}\,d\eta_1 \,d^3 r~.
\end{equation}
Using the vector identity 
\begin{equation}
	\hat{\boldsymbol{s}} \cdot\left[\boldsymbol{\nabla}V^{I} \times \delta\boldsymbol{A}\right]=\boldsymbol{\nabla}V^{I} \cdot\left[\delta\boldsymbol{A} \times \hat{\boldsymbol{s}}\right]=\delta\boldsymbol{A} \cdot\left[\hat{\boldsymbol{s}} \times \boldsymbol{\nabla}V^{I}\right]
\end{equation}
we obtain
\begin{equation}
	\delta {H}_{c}=g_{\textrm{e}}\frac{ne}{c^2}\sum_{I=1}^{N} \int \left \{  \Psi^{*} \mathcal{K}^2\left[\hat{\boldsymbol{s}} \times \boldsymbol{\nabla}V^{I}\right]\Psi\,d\boldsymbol{X}_{1}\,d\eta_1  \right \}\cdot\delta\boldsymbol{A}\,d^3 r
\end{equation}
that enables us to achieve as contribution to the total induced electron current density, using relation (\ref{classic H variation}), the expression
\begin{equation}
\begin{split}
	\boldsymbol{J}(\boldsymbol{r})=&-g_{\textrm{e}}\frac{e^2}{c^2}\mathcal{K}^2(\boldsymbol{r})\sum_{I=1}^{N} \boldsymbol{Q}(\boldsymbol{r})\times\boldsymbol{\nabla}V^{I}\\=&-g_{\textrm{e}}\frac{e^2}{c^2}\mathcal{K}^2(\boldsymbol{r}) \boldsymbol{Q}(\boldsymbol{r})\times\boldsymbol{\nabla}V\\=&+g_{\textrm{e}}\frac{e^2}{c^2}\mathcal{K}^2(\boldsymbol{r}) \boldsymbol{\nabla}V\times\boldsymbol{Q}(\boldsymbol{r})~.
\label{spin-orbit coupling current}
\end{split}
\end{equation}
While this expression is regularized at the nuclear positions by employing 
finite nuclear charge distributions instead of point charges, it vanishes 
entirely when added to the previously derived magnetization contribution. 
This marks a crucial difference between ZORA and the Breit--Pauli approach.
\section{Continuity of ZORA Spin Current Density}
\label{continuity}
In this section, the continuity condition on spin contributions to the total magnetically induced current density, 
Eq.~\eqref{ZORAtttcurrent}, is analyzed. Let us start by examining the spin Zeeman contribution to the current density, Eq.~\eqref{magnet-curr}. 
Its divergence is given by:
\begin{equation}
	\boldsymbol{\nabla} \cdot \boldsymbol{J}(\boldsymbol{r}) = -g_{\textrm{e}}\mu_{B}\frac{2m_{\textrm{e}}}{\hbar} \boldsymbol{\nabla} \cdot \left[ \mathcal{K}(\boldsymbol{r}) \boldsymbol{\nabla} \times \boldsymbol{Q}(\boldsymbol{r}) \right]
\end{equation}
By applying the vector identity 
\begin{equation}
\boldsymbol{\nabla} \cdot (f\boldsymbol{A}) = \boldsymbol{\nabla} f \cdot \boldsymbol{A} + f (\boldsymbol{\nabla} \cdot \boldsymbol{A})
\label{vectdivide}	
\end{equation}
with $f= \mathcal{K}(\boldsymbol{r})$ and $\boldsymbol{A}=\boldsymbol{\nabla} \times \boldsymbol{Q}(\boldsymbol{r})$, 
we obtain
\begin{equation}
\begin{split}
	\boldsymbol{\nabla} \cdot \boldsymbol{J}(\boldsymbol{r}) = &-g_{\textrm{e}}\mu_{B}\frac{2m_{\textrm{e}}}{\hbar} \left\{ \left[ \boldsymbol{\nabla} \mathcal{K}(\boldsymbol{r}) \right] \cdot \left[ \boldsymbol{\nabla} \times \boldsymbol{Q}(\boldsymbol{r}) \right] \right\}\\&-g_{\textrm{e}}\mu_{B}\frac{2m_{\textrm{e}}}{\hbar}\left\{ \mathcal{K}(\boldsymbol{r}) \boldsymbol{\nabla} \cdot \left[ \boldsymbol{\nabla} \times \boldsymbol{Q}(\boldsymbol{r}) \right] \right\}
\end{split}
\end{equation}
The first term vanishes in atomic central potentials, as can be seen by inspecting Eq.~\eqref{Kderiv}, due to the local orthogonality between the radial electric field generated by the nuclear charges and the transverse spin magnetization electron current density vector field. The last term is the divergence of a curl \cite{summa_molecular_2024}.
\section{Statistical Mechanics of Spin Expectation Value}
\label{TERMOS}
\noindent
The most ``classical'' general expression for the spin expectation value is~\cite{lewis2020thermodynamics}
\begin{equation}
	\begin{split}
		\langle S_{\text{op},\beta}\rangle=&-\frac{(2S+1)}{2}\,\coth\left[\frac{(2S+1)g_{\textrm{e}}\mu_{B}B_{\beta}}{2k_B T}\right]\\&+\frac{1}{2}\,\coth\left[\frac{g_{\textrm{e}}\mu_{B}B_{\beta}}{2k_B T}\right]
	\end{split}
	\label{spinstatcomp}
\end{equation}
This can be written compactly as
\begin{equation}
	\langle S_{\text{op},\beta}\rangle=-\left[k\,\coth\left(kx\right)-\frac{1}{2}\,\coth\left(\frac{x}{2}\right)\right]
\end{equation}
where $k=\frac{2S+1}{2}$ and $x=\frac{g_{\textrm{e}}\mu_{B}B_{\beta}}{k_B T}$. Using the series expansion
\begin{equation}
	\coth\left(x\right)\simeq\frac{1}{x}+\frac{x}{3}-\frac{x^3}{45}+\cdots
\end{equation}
and retaining only the first two terms, we obtain
\begin{equation}
	\langle S_{\text{op},\beta}\rangle \simeq-\left[\frac{k^2x}{3}-\frac{x}{12}\right] 
	= -S(S+1)\frac{g_{\textrm{e}}\mu_{B}B_{\beta}}{3k_B T}
\end{equation}

This recovers the familiar high-temperature (or weak-field) linear approximation used in the main text during derivations of spin contributions to nuclear magnetic shielding and magnetizability tensors. 

The complete expression (\ref{spinstatcomp}) provides a general framework to obtain the spin contribution to both shielding and magnetizability at any temperature $T$. By defining the derivative of the expectation value of the spin operator with respect to the magnetic field as:
\begin{equation}
	\begin{split}
		\frac{\partial\langle S_{\text{op},\beta} \rangle}{\partial B_\beta} &=\frac{g_{\textrm{e}}\mu_B}{4k_B T}(2S+1)^2\text{csch}^2\left[\frac{(2S+1)g_{\textrm{e}}\mu_B B_{\beta}}{2k_B T}\right]\\&-  \frac{g_{\textrm{e}}\mu_B}{4k_B T}\text{csch}^2\left[\frac{g_{\textrm{e}}\mu_B B_{\beta}}{2k_B T}\right]
	\end{split}
\end{equation}
it can be shown that:
\begin{equation}
	\lim_{T\rightarrow 0\text{ K}} \frac{\partial\langle S_{\text{op},\beta} \rangle}{\partial B_\beta} = 0
\end{equation}

Consequently, the spin contributions to shielding and magnetizability vanish at $0\text{ K}$. This behavior is physically consistent with the reaching of the saturation limit, as described by the Brillouin-Langevin theory of paramagnetism~\cite{Kittel2004,Atkins2023}. At absolute zero, the system is locked in its ground state (maximal alignment with the field), and the differential spin magnetizability vanishes as the system can no longer be further polarized. 

A similar vanishing behavior is obtained in the limit $T \rightarrow \infty$, albeit for a different physical reason: at infinite temperature, all spin states become equally populated due to thermal agitation, causing the net magnetization to vanish and making the system's response to the external field negligible. 

It is worth noting that while electrons are intrinsically fermions, the use of Maxwell-Boltzmann (MB) statistics for open-shell molecules is justified by the fact that these systems typically operate in the non-degenerate regime. In molecular gases or diluted radical solutions, the spatial separation between molecules ensures that the electronic wavefunctions do not overlap, effectively rendering the spin centers distinguishable and non-interacting. 

Consequently, the derivation based on the Brillouin function introduced here, and specifically the expression (\ref{spinstatcomp}), stems directly from the application of MB statistics and remains the standard framework for molecular paramagnetism. This theoretical approach is rigorously valid only within the classical limits of the MB distribution, namely in the high-temperature regime for dilute systems or in the presence of a weak magnetic field. 

In contrast, in systems characterized by significant electronic degeneracy, the fermionic nature of the spin carriers would necessitate the use of Fermi-Dirac statistics to provide a consistent description of shieldings and magnetizabilities across the entire temperature range.

\section{Effective Electrostatic Potential $V(\boldsymbol{r})$}
\label{veff}
The effective electrostatic potential introduced in the main text is evaluated, in our approach, as
\begin{equation}
	\begin{split}
		V(\boldsymbol{r})=&-\sum_{I=1}^{N}\sum_{i=1}^{k}c_i\frac{\textrm{erf}(\sqrt{\alpha_i}\left|\boldsymbol{r}-\boldsymbol{R}_I\right|)}{\left|\boldsymbol{r}-\boldsymbol{R}_I\right|}\\&-\sum_{I}^{N}\frac{Z_I}{\left|\boldsymbol{r}-\boldsymbol{R}_I\right|}\textrm{erf}(\sqrt{\zeta_I}\left|\boldsymbol{r}-\boldsymbol{R}_I\right|)
	\end{split}
\end{equation}
by using relativistic data  provided by Lehtola et al. in Ref. [\citenum{lehtola_efficient_2020}] (with parameters $c_i$, $\alpha_i$ and the index $k$ that depends from the atomic dataset) and $\zeta_I$ obtained using the recipe of Ref. [\citenum{visscher_diracfock_1997}] 
\begin{equation}
	\zeta_I = 1.5 \times 10^{10} \left( \frac{a_0}{0.836 A^{1/3} + 0.570}\right)^2
\end{equation}
where $a_0$ is Bohr's radius in Angstrom and  $A$ is  the atomic mass number, used to account for the finite size of the Gaussian nuclei here considered. Note that the previous potential removes the divergence at nuclear position, indeed for an atom
\begin{equation}
	V_I(\boldsymbol{R}_I)=-\sum_{i=1}^{k}2c_i\frac{\sqrt{\alpha_i}}{\sqrt{\pi}}-2Z_I\frac{\sqrt{\zeta_I}}{\sqrt{\pi}}
	\label{potnuc}
\end{equation}
Its gradient is computed as follows
\begin{widetext}
	\begin{equation}
		\begin{split}
			\boldsymbol{\nabla}V(\boldsymbol{r}) = & \sum_{I=1}^{N} \sum_{i=1}^{k} c_{i} \frac{\boldsymbol{r} - \boldsymbol{R}_I}{|\boldsymbol{r} - \boldsymbol{R}_I|}\left(\frac{\text{erf}(\sqrt{\alpha_i} |\boldsymbol{r} - \boldsymbol{R}_I|)}{|\boldsymbol{r} - \boldsymbol{R}_I|^2}-\frac{2\sqrt{\alpha_i}}{\sqrt{\pi}} \frac{e^{-\alpha_i |\boldsymbol{r} - \boldsymbol{R}_I|^2}}{|\boldsymbol{r} - \boldsymbol{R}_I|} \right) \\
			& + \sum_{I=1}^{N} Z_{I} \frac{\boldsymbol{r} - \boldsymbol{R}_I}{|\boldsymbol{r} - \boldsymbol{R}_I|}\left( \frac{\text{erf}(\sqrt{\zeta_I} |\boldsymbol{r} - \boldsymbol{R}_I|)}{|\boldsymbol{r} - \boldsymbol{R}_I|^2} 
			- \frac{2\sqrt{\zeta_I}}{\sqrt{\pi}} \frac{e^{-\zeta_I |\boldsymbol{r} - \boldsymbol{R}_I|^2}}{|\boldsymbol{r} - \boldsymbol{R}_I|} \right)
		\end{split}
	\end{equation}
\end{widetext}
For nuclear positions, as illustrated by inspecting equation (\ref{potnuc}), we get 
\begin{equation}
	\boldsymbol{\nabla}V_I(\boldsymbol{R}_I)=0
\end{equation}

\bibliography{bibliostcu}

@string{jacs = "J. Am. Chem. Soc."}

@article{blugel_hyperfine_1987,
  title = {Hyperfine fields of 3<i>d</i>and 4<i>d</i>impurities in nickel},
volume = {35},
ISSN = {0163-1829},
url = {http://dx.doi.org/10.1103/PhysRevB.35.3271},
DOI = {10.1103/physrevb.35.3271},
number = {7},
journal = {Physical Review B},
publisher = {American Physical Society (APS)},
author = {Bl\"{u}gel,  S. and Akai,  H. and Zeller,  R. and Dederichs,  P. H.},
year = {1987},
month = Mar,
pages = {3271–3283}
}

@article{luo_theoretical_2021,
  title = {Theoretical study on radii of neutral atoms and singly charged negative ions},
volume = {138},
ISSN = {0092-640X},
url = {http://dx.doi.org/10.1016/j.adt.2020.101392},
DOI = {10.1016/j.adt.2020.101392},
journal = {Atomic Data and Nuclear Data Tables},
publisher = {Elsevier BV},
author = {Luo,  Mingmin and Min,  Guangxin and Guo,  Guannan and Zhang,  Xuemei},
year = {2021},
month = Mar,
pages = {101392}
}

@article{treutler_efficient_1995,
  title = {Efficient molecular numerical integration schemes},
volume = {102},
ISSN = {1089-7690},
url = {http://dx.doi.org/10.1063/1.469408},
DOI = {10.1063/1.469408},
number = {1},
journal = {The Journal of Chemical Physics},
publisher = {AIP Publishing},
author = {Treutler,  Oliver and Ahlrichs,  Reinhart},
year = {1995},
month = Jan,
pages = {346–354}
}

@article{rouf_relativistic_2017,
  title = {Relativistic Approximations to Paramagnetic NMR Chemical Shift and Shielding Anisotropy in Transition Metal Systems},
volume = {13},
ISSN = {1549-9626},
url = {http://dx.doi.org/10.1021/acs.jctc.7b00168},
DOI = {10.1021/acs.jctc.7b00168},
number = {8},
journal = {Journal of Chemical Theory and Computation},
publisher = {American Chemical Society (ACS)},
author = {Rouf,  Syed Awais and Mareš,  Jiří and Vaara,  Juha},
year = {2017},
pages = {3731–3745}
}

@article{speelman_nmr_2025,
  title = {NMR chemical shielding for solid-state systems using spin–orbit coupled ZORA GIPAW},
volume = {163},
ISSN = {1089-7690},
url = {http://dx.doi.org/10.1063/5.0278794},
DOI = {10.1063/5.0278794},
number = {10},
journal = {The Journal of Chemical Physics},
publisher = {AIP Publishing},
author = {Speelman,  T. and Huebsch,  M.-T. and Havenith,  R. W. A. and Marsman,  M. and de Wijs,  G. A.},
year = {2025},
}

@book{Kittel2004,
	author = {Kittel, Charles},
	description = {Amazon.com: Introduction to Solid State Physics (9780471415268): Charles Kittel: Books},
	dewey = {530.41},
	ean = {9780471415268},
	edition = 8,
	isbn = {9780471415268},
	publisher = {Wiley},
	title = {Introduction to Solid State Physics},
	url = {http://www.amazon.com/Introduction-Solid-Physics-Charles-Kittel/dp/047141526X/ref=dp_ob_title_bk},
	year = 2004
}

@incollection{autschbach_nmr_2015,
title = {Chapter One - NMR Calculations for Paramagnetic Molecules and Metal Complexes},
editor = {David A. Dixon},
series = {Annual Reports in Computational Chemistry},
volume = {11},
pages = {3-36},
year = {2015},
doi = {https://doi.org/10.1016/bs.arcc.2015.09.006},
author = {Jochen Autschbach},
ISBN = {9780444637109},
ISSN = {1574-1400},
url = {http://dx.doi.org/10.1016/bs.arcc.2015.09.006},
booktitle = {Annual Reports in Computational Chemistry},
publisher = {Elsevier},
}

@article{martin_temperature_2015,
 title = {Temperature dependence of contact and dipolar NMR chemical shifts in paramagnetic molecules},
volume = {142},
ISSN = {1089-7690},
url = {http://dx.doi.org/10.1063/1.4906318},
DOI = {10.1063/1.4906318},
number = {5},
journal = {The Journal of Chemical Physics},
publisher = {AIP Publishing},
author = {Martin,  Bob and Autschbach,  Jochen},
year = {2015},
month = Feb 
}

@article{van_den_heuvel_nmr_2013,
  title = {NMR chemical shift as analytical derivative of the Helmholtz free energy},
volume = {138},
ISSN = {1089-7690},
url = {http://dx.doi.org/10.1063/1.4789398},
DOI = {10.1063/1.4789398},
number = {5},
journal = {The Journal of Chemical Physics},
publisher = {AIP Publishing},
author = {Van den Heuvel,  Willem and Soncini,  Alessandro},
year = {2013},
}

@article{franzke_reducing_2023,
  title = {Reducing Exact Two-Component Theory for NMR Couplings to a One-Component Approach: Efficiency and Accuracy},
volume = {19},
ISSN = {1549-9626},
url = {http://dx.doi.org/10.1021/acs.jctc.2c01248},
DOI = {10.1021/acs.jctc.2c01248},
number = {7},
journal = {Journal of Chemical Theory and Computation},
publisher = {American Chemical Society (ACS)},
author = {Franzke,  Yannick J.},
year = {2023},
pages = {2010–2028}
}

@article{kutzelnigg_relativistic_2009,
 title = {Relativistic theory of nuclear magnetic resonance parameters in a Gaussian basis representation},
volume = {131},
ISSN = {1089-7690},
url = {http://dx.doi.org/10.1063/1.3185400},
DOI = {10.1063/1.3185400},
number = {4},
journal = {The Journal of Chemical Physics},
publisher = {AIP Publishing},
author = {Kutzelnigg,  Werner and Liu,  Wenjian},
year = {2009},
}

@article{kutzelnigg_quasirelativistic_2005,
  title = {Quasirelativistic theory equivalent to fully relativistic theory},
volume = {123},
ISSN = {1089-7690},
url = {http://dx.doi.org/10.1063/1.2137315},
DOI = {10.1063/1.2137315},
number = {24},
journal = {The Journal of Chemical Physics},
publisher = {AIP Publishing},
author = {Kutzelnigg,  Werner and Liu,  Wenjian},
year = {2005},
}

@article{ilias_infinite-order_2007,
  title = {An infinite-order two-component relativistic Hamiltonian by a simple one-step transformation},
volume = {126},
ISSN = {1089-7690},
url = {http://dx.doi.org/10.1063/1.2436882},
DOI = {10.1063/1.2436882},
number = {6},
journal = {The Journal of Chemical Physics},
publisher = {AIP Publishing},
author = {Iliaš,  Miroslav and Saue,  Trond},
year = {2007},
}

@article{hess_relativistic_1986,
  title = {Relativistic electronic-structure calculations employing a two-component no-pair formalism with external-field projection operators},
volume = {33},
ISSN = {0556-2791},
url = {http://dx.doi.org/10.1103/PhysRevA.33.3742},
DOI = {10.1103/physreva.33.3742},
number = {6},
journal = {Physical Review A},
publisher = {American Physical Society (APS)},
author = {Hess,  Bernd A.},
year = {1986},
pages = {3742–3748}
}

@article{douglas_quantum_1974,
 title = {Quantum electrodynamical corrections to the fine structure of helium},
volume = {82},
ISSN = {0003-4916},
url = {http://dx.doi.org/10.1016/0003-4916(74)90333-9},
DOI = {10.1016/0003-4916(74)90333-9},
number = {1},
journal = {Annals of Physics},
publisher = {Elsevier BV},
author = {Douglas,  Marvin and Kroll,  Norman M},
year = {1974},
pages = {89–155}
}

@article{reiher_exact_2004,
  title = {Exact decoupling of the Dirac Hamiltonian. II. The generalized Douglas–Kroll–Hess transformation up to arbitrary order},
volume = {121},
ISSN = {1089-7690},
url = {http://dx.doi.org/10.1063/1.1818681},
DOI = {10.1063/1.1818681},
number = {22},
journal = {The Journal of Chemical Physics},
publisher = {AIP Publishing},
author = {Reiher,  Markus and Wolf,  Alexander},
year = {2004},
pages = {10945–10956}
}

@article{rinkevicius_calculations_2003,
  title = {Calculations of nuclear magnetic shielding in paramagnetic molecules},
volume = {118},
ISSN = {1089-7690},
url = {http://dx.doi.org/10.1063/1.1535904},
DOI = {10.1063/1.1535904},
number = {6},
journal = {The Journal of Chemical Physics},
publisher = {AIP Publishing},
author = {Rinkevicius,  Zilvinas and Vaara,  Juha and Telyatnyk,  Lyudmyla and Vahtras,  Olav},
year = {2003},
pages = {2550–2561}
}

@article{martin_erratum_2016,
  title = {Erratum: “Temperature dependence of contact and dipolar NMR chemical shifts in paramagnetic molecules” [J. Chem. Phys. 142,  054108 (2015)]},
volume = {145},
ISSN = {1089-7690},
url = {http://dx.doi.org/10.1063/1.4959030},
DOI = {10.1063/1.4959030},
number = {4},
journal = {The Journal of Chemical Physics},
publisher = {AIP Publishing},
author = {Martin,  Bob and Autschbach,  Jochen},
year = {2016},
}

@book{Atkins2023,
	author    = {Atkins, Peter and de Paula, Julio and Keeler, James},
	title     = {Atkins' Physical Chemistry},
	publisher = {Oxford University Press},
	year      = {2023},
	edition   = {12th},
	address   = {Oxford, United Kingdom},
	isbn      = {978-0198847816}
}

@article{hong_comparison_2001,
  title = {A comparison of scalar-relativistic ZORA and DKH density functional schemes: monohydrides,  monooxides and monofluorides of La,  Lu,  Ac and Lr},
volume = {334},
ISSN = {0009-2614},
url = {http://dx.doi.org/10.1016/S0009-2614(00)01430-5},
DOI = {10.1016/s0009-2614(00)01430-5},
number = {4-6},
journal = {Chemical Physics Letters},
publisher = {Elsevier BV},
author = {Hong,  Gongyi and Dolg,  Michael and Li,  Lemin},
year = {2001},
month = Feb,
pages = {396–402}
}

@article{nakajima_douglaskrollhess_2012,
  title = {The Douglas–Kroll–Hess Approach},
volume = {112},
ISSN = {1520-6890},
url = {http://dx.doi.org/10.1021/cr200040s},
DOI = {10.1021/cr200040s},
number = {1},
journal = {Chemical Reviews},
publisher = {American Chemical Society (ACS)},
author = {Nakajima,  Takahito and Hirao,  Kimihiko},
year = {2011},
pages = {385–402}
}

@article{van_lenthe_relativistic_1994,
  title = {Relativistic total energy using regular approximations},
volume = {101},
ISSN = {1089-7690},
url = {http://dx.doi.org/10.1063/1.467943},
DOI = {10.1063/1.467943},
number = {11},
journal = {The Journal of Chemical Physics},
publisher = {AIP Publishing},
author = {van Lenthe,  Erik and Baerends,  Evert-Jan and Snijders,  Jaap G.},
year = {1994},
month = Dec,
pages = {9783–9792}
}

@article{van_lenthe_geometry_1999,
  title = {Geometry optimizations in the zero order regular approximation for relativistic effects},
volume = {110},
ISSN = {1089-7690},
url = {http://dx.doi.org/10.1063/1.478813},
DOI = {10.1063/1.478813},
number = {18},
journal = {The Journal of Chemical Physics},
publisher = {AIP Publishing},
author = {van Lenthe,  Erik and Ehlers,  Andreas and Baerends,  Evert-Jan},
year = {1999},
month = May,
pages = {8943–8953}
}

@book{dyall_introduction_2007,
  title = {Introduction to Relativistic Quantum Chemistry},
ISBN = {9780197561744},
url = {http://dx.doi.org/10.1093/oso/9780195140866.001.0001},
DOI = {10.1093/oso/9780195140866.001.0001},
publisher = {Oxford University Press},
author = {Dyall,  Kenneth G. and Faegri,  Knut},
year = {2007},
}

@article{lehtola_efficient_2020,
  title = {Efficient implementation of the superposition of atomic potentials initial guess for electronic structure calculations in Gaussian basis sets},
volume = {152},
ISSN = {1089-7690},
url = {http://dx.doi.org/10.1063/5.0004046},
DOI = {10.1063/5.0004046},
number = {14},
journal = {The Journal of Chemical Physics},
publisher = {AIP Publishing},
author = {Lehtola,  Susi and Visscher,  Lucas and Engel,  Eberhard},
year = {2020},
month = Apr 
}

@article{van_wullen_molecular_1998,
 title = {Molecular density functional calculations in the regular relativistic approximation: Method,  application to coinage metal diatomics,  hydrides,  fluorides and chlorides,  and comparison with first-order relativistic calculations},
volume = {109},
ISSN = {1089-7690},
url = {http://dx.doi.org/10.1063/1.476576},
DOI = {10.1063/1.476576},
number = {2},
journal = {The Journal of Chemical Physics},
publisher = {AIP Publishing},
author = {van W\"{u}llen,  Christoph},
year = {1998},
pages = {392–399}
}

@incollection{autschbach_chapter_2009,
  title = {Chapter 1 Relativistic Computations of NMR Parameters from First Principles: Theory and Applications},
ISBN = {9780123750587},
ISSN = {0066-4103},
url = {http://dx.doi.org/10.1016/S0066-4103(09)06701-5},
DOI = {10.1016/s0066-4103(09)06701-5},
booktitle = {Annual Reports on NMR Spectroscopy},
publisher = {Elsevier},
author = {Autschbach,  J. and Zheng,  S.},
year = {2009},
pages = {1–95}
}

@article{romaniello_relativistic_2007,
  title = {Relativistic two-component formulation of time-dependent current-density functional theory: Application to the linear response of solids},
volume = {127},
ISSN = {1089-7690},
url = {http://dx.doi.org/10.1063/1.2780146},
DOI = {10.1063/1.2780146},
number = {17},
journal = {The Journal of Chemical Physics},
publisher = {AIP Publishing},
author = {Romaniello,  Pina and de Boeij,  Paul L.},
year = {2007},
month = Nov 
}

@article{visscher_diracfock_1997,
  title = {DIRAC–FOCK ATOMIC ELECTRONIC STRUCTURE CALCULATIONS USING DIFFERENT NUCLEAR CHARGE DISTRIBUTIONS},
volume = {67},
ISSN = {0092-640X},
url = {http://dx.doi.org/10.1006/adnd.1997.0751},
DOI = {10.1006/adnd.1997.0751},
number = {2},
journal = {Atomic Data and Nuclear Data Tables},
publisher = {Elsevier BV},
author = {Visscher,  Lucas and Dyall,  Kennetg G.},
year = {1997},
month = Nov,
pages = {207–224}
}

@book{bjorken1964relativistic,
 author = "Bjorken, James D. and Drell, Sidney D.",
title = "{Relativistic Quantum Mechanics}",
isbn = "978-0-07-005493-6",
publisher = "McGraw-Hill",
address = "New York",
series = "International Series In Pure and Applied Physics",
year = "1965"
}

@book{CRC_Handbook_2024,
	editor    = {John R. Rumble},
	title     = {CRC Handbook of Chemistry and Physics},
	edition   = {105th},
	publisher = {CRC Press/Taylor \& Francis},
	address   = {Boca Raton, FL},
	year      = {2024},
	isbn      = {978-1032770222}
}

@article{Havens1932,
  title = {The Magnetic Susceptibility of Nitrogen Dioxide},
volume = {41},
ISSN = {0031-899X},
url = {http://dx.doi.org/10.1103/PhysRev.41.337},
DOI = {10.1103/physrev.41.337},
number = {3},
journal = {Physical Review},
publisher = {American Physical Society (APS)},
author = {Havens,  Glenn G.},
year = {1932},
month = Aug,
pages = {337–344}
}

@article{franzke_paramagnetic_2024,
  title = {Paramagnetic Nuclear Magnetic Resonance Shifts for Triplet Systems and Beyond with Modern Relativistic Density Functional Methods},
volume = {128},
ISSN = {1520-5215},
url = {http://dx.doi.org/10.1021/acs.jpca.3c07093},
DOI = {10.1021/acs.jpca.3c07093},
number = {3},
journal = {The Journal of Physical Chemistry A},
publisher = {American Chemical Society (ACS)},
author = {Franzke,  Yannick J. and Bruder,  Florian and Gillhuber,  Sebastian and Holzer,  Christof and Weigend,  Florian},
year = {2024},
month = Jan,
pages = {670–686}
}

@article{stevens_perturbed_1963,
  title = {Perturbed Hartree—Fock Calculations. I. Magnetic Susceptibility and Shielding in the LiH Molecule},
volume = {38},
ISSN = {1089-7690},
url = {http://dx.doi.org/10.1063/1.1733693},
DOI = {10.1063/1.1733693},
number = {2},
journal = {The Journal of Chemical Physics},
publisher = {AIP Publishing},
author = {Stevens,  R. M. and Pitzer,  Russel M. and Lipscomb,  William N.},
year = {1963},
month = Jan,
pages = {550–560}
}

@article{kern_magnetic_1962,
  title = {Magnetic Shielding in Some Diatomic Molecules},
volume = {37},
ISSN = {1089-7690},
url = {http://dx.doi.org/10.1063/1.1701314},
DOI = {10.1063/1.1701314},
number = {2},
journal = {The Journal of Chemical Physics},
publisher = {AIP Publishing},
author = {Kern,  C. William and Lipscomb,  William N.},
year = {1962},
pages = {260–266}
}

@book{lewis2020thermodynamics,
	title={Thermodynamics},
	author={Lewis, G.N. and Randall, M. and Pitzer, K.S. and Brewer, L.},
	isbn={9780486842745},
	series={Dover Books on Chemistry},
	year={2020},
	publisher={Dover Publications}
}

@book{summa_molecular_2024,
  title = {Molecular Properties via Induced Current Densities},
ISBN = {9783031601590},
ISSN = {2190-5061},
url = {http://dx.doi.org/10.1007/978-3-031-60159-0},
DOI = {10.1007/978-3-031-60159-0},
journal = {Springer Theses},
publisher = {Springer Nature Switzerland},
author = {Summa,  Francesco Ferdinando},
year = {2024}
}

@article{zanasi_ring_1995,
	title = {Ring currents and magnetisability in {C60}},
	volume = {238},
	issn = {00092614},
	url = {https://linkinghub.elsevier.com/retrieve/pii/0009261495004379},
	doi = {10.1016/0009-2614(95)00437-9},
	number = {4-6},
	urldate = {2022-03-03},
	journal = {Chem. Phys. Lett.},
	author = {Zanasi, R. and Fowler, P.W.},
	month = jun,
	year = {1995},
	pages = {270--280},
}

@article{monaco_atomic_2020,
	title = {Atomic size adjusted calculation of the magnetically induced current density},
	volume = {745},
	issn = {00092614},
	url = {https://linkinghub.elsevier.com/retrieve/pii/S0009261420301962},
	doi = {10.1016/j.cplett.2020.137281},
	urldate = {2022-10-18},
	journal = {Chem. Phys. Lett.},
	author = {Monaco, Guglielmo and Summa, Francesco F. and Zanasi, Riccardo},
	month = apr,
	year = {2020},
	pages = {137281},
}

@book{epstein_variation_1974,
	address = {New York},
	series = {Physical chemistry, a series of monographs},
	title = {The variation method in quantum chemistry},
	isbn = {978-0-12-240550-1},
	number = {v. 33},
	publisher = {Academic Press},
	author = {Epstein, Saul T.},
	year = {1974},
}

@article{becke_numerical_1988,
  title = {Numerical solution of Poisson’s equation in polyatomic molecules},
volume = {89},
ISSN = {1089-7690},
url = {http://dx.doi.org/10.1063/1.455005},
DOI = {10.1063/1.455005},
number = {5},
journal = {The Journal of Chemical Physics},
publisher = {AIP Publishing},
author = {Becke,  A. D. and Dickson,  R. M.},
year = {1988},
pages = {2993–2997}
}

@book{greiner_field_1996,
  title = {Field Quantization},
ISBN = {9783642614859},
url = {http://dx.doi.org/10.1007/978-3-642-61485-9},
DOI = {10.1007/978-3-642-61485-9},
publisher = {Springer Berlin Heidelberg},
author = {Greiner,  Walter and Reinhardt,  Joachim},
year = {1996}
}

@article{lazzeretti_methods_2012,
  title = {Methods of continuous translation of the origin of the current density revisited},
volume = {131},
ISSN = {1432-2234},
url = {http://dx.doi.org/10.1007/s00214-012-1222-y},
DOI = {10.1007/s00214-012-1222-y},
number = {5},
journal = {Theoretical Chemistry Accounts},
publisher = {Springer Science and Business Media LLC},
author = {Lazzeretti,  P.},
year = {2012},
month = Apr 
}

@article{Lazzeretti2012erratum,
	title = {Erratum to: Methods of continuous translation of the origin of the current density revisited},
	volume = {132},
	ISSN = {1432-2234},
	url = {http://dx.doi.org/10.1007/s00214-012-1317-5},
	DOI = {10.1007/s00214-012-1317-5},
	number = {2},
	journal = {Theoretical Chemistry Accounts},
	publisher = {Springer Science and Business Media LLC},
	author = {Lazzeretti,  P.},
	year = {2012},
	month = Dec 
}

@article{sundholm_current_2021,
	title = {Current density and molecular magnetic properties},
	volume = {57},
	issn = {1359-7345, 1364-548X},
	url = {http://xlink.rsc.org/?DOI=D1CC03350F},
	doi = {10.1039/D1CC03350F},
	number = {93},
	urldate = {2022-03-18},
	journal = {Chem. Commun.},
	author = {Sundholm, Dage and Dimitrova, Maria and Berger, Raphael J. F.},
	year = {2021},
	pages = {12362--12378},
}

@misc{g16,
	author={M. J. Frisch and G. W. Trucks and H. B. Schlegel and G. E. Scuseria and M. A. Robb and J. R. Cheeseman and G. Scalmani and V. Barone and G. A. Petersson and H. Nakatsuji and X. Li and M. Caricato and A. V. Marenich and J. Bloino and B. G. Janesko and R. Gomperts and B. Mennucci and H. P. Hratchian and J. V. Ortiz and A. F. Izmaylov and J. L. Sonnenberg and D. Williams-Young and F. Ding and F. Lipparini and F. Egidi and J. Goings and B. Peng and A. Petrone and T. Henderson and D. Ranasinghe and V. G. Zakrzewski and J. Gao and N. Rega and G. Zheng and W. Liang and M. Hada and M. Ehara and K. Toyota and R. Fukuda and J. Hasegawa and M. Ishida and T. Nakajima and Y. Honda and O. Kitao and H. Nakai and T. Vreven and K. Throssell and Montgomery, {Jr.}, J. A. and J. E. Peralta and F. Ogliaro and M. J. Bearpark and J. J. Heyd and E. N. Brothers and K. N. Kudin and V. N. Staroverov and T. A. Keith and R. Kobayashi and J. Normand and K. Raghavachari and A. P. Rendell and J. C. Burant and S. S. Iyengar and J. Tomasi and M. Cossi and J. M. Millam and M. Klene and C. Adamo and R. Cammi and J. W. Ochterski and R. L. Martin and K. Morokuma and O. Farkas and J. B. Foresman and D. J. Fox},
	title={Gaussian16 {R}evision {C}.01},
	year={2016},
	note={Gaussian Inc. Wallingford CT}
}

@book{landau_quantum_2007,
	address = {Singapore},
	edition = {3. ed., rev. and enl., authorized Engl. reprint ed},
	series = {Course of theoretical physics / by {L}. {D}. {Landau} and {E}. {M}. {Lifshitz}},
	title = {Quantum mechanics: non-relativistic theory},
	isbn = {978-0-7506-3539-4 978-981-272-088-7 978-7-5062-4257-8},
	shorttitle = {Quantum mechanics},
	number = {Vol. 3},
	publisher = {Elsevier [u.a.]},
	author = {Landau, Lev Davidovič and Lifšic, Evgenij M. and Landau, Lev Davidovič},
	year = {2007},
}

@article{gordon_strom_1928,
  title = {Der Strom der Diracschen Elektronentheorie},
volume = {50},
ISSN = {1434-601X},
url = {http://dx.doi.org/10.1007/BF01327881},
DOI = {10.1007/bf01327881},
number = {9-10},
journal = {Zeitschrift f\"{u}r Physik},
publisher = {Springer Science and Business Media LLC},
author = {Gordon,  W.},
year = {1928},
pages = {630–632}
}

@article{Blasco2026,
  author  = {Blasco, Daniel and Novotn{\'y}, Jan and Asher, James R. and Berger, Raphael J. F. and Komorovsk{\'y}, Stanislav and Marek, Radek},
  title   = {Link between Spin-Orbit Relativity and Magnetically Induced Current Densities in Heavy-Atom Hydrides: trans-Ligand Influence},
  journal = {JACS Au},
  year    = {2026},
  doi     = {10.1021/jacsau.6c00346},
  url     = {https://doi.org/10.1021/jacsau.6c00346},
  note    = {In press}
}

@article{soncini_charge_2007,
  title = {{Charge and Spin Currents in Open-Shell Molecules: A Unified Description of NMR and EPR Observables}},
volume = {3},
ISSN = {1549-9626},
url = {http://dx.doi.org/10.1021/ct700169h},
DOI = {10.1021/ct700169h},
number = {6},
journal = {Journal of Chemical Theory and Computation},
publisher = {American Chemical Society (ACS)},
author = {Soncini,  Alessandro},
year = {2007},
month = Oct,
pages = {2243–2257}
}

@article{jameson_nuclear_1979,
  title = {Nuclear magnetic shielding density},
volume = {83},
ISSN = {1541-5740},
url = {http://dx.doi.org/10.1021/j100489a011},
DOI = {10.1021/j100489a011},
number = {26},
journal = {The Journal of Physical Chemistry},
publisher = {American Chemical Society (ACS)},
author = {Jameson,  Cynthia J. and Buckingham,  A. D.},
year = {1979},
month = Dec,
pages = {3366–3371}
}

@book{mcweeny_spins_1970,
	address = {New York},
	series = {Current chemical concepts},
	title = {Spins in chemistry},
	isbn = {978-0-12-486750-5},
	publisher = {Academic Press},
	author = {McWeeny, R.},
	collaborator = {Polytechnic Institute of Brooklyn},
	year = {1970},
}

@book{mcweeny_methods_1992,
	address = {London},
	edition = {2. Aufl},
	series = {Theoretical chemistry},
	title = {Methods of molecular quantum mechanics},
	isbn = {978-0-12-486552-5},
	publisher = {Academic Press},
	author = {McWeeny, Roy},
	year = {1992},
}

@article{keith_calculation_1992,
	title = {Calculation of magnetic response properties using atoms in molecules},
	volume = {194},
	issn = {00092614},
	url = {https://linkinghub.elsevier.com/retrieve/pii/000926149285733Q},
	doi = {10.1016/0009-2614(92)85733-Q},
	number = {1-2},
	urldate = {2022-06-02},
	journal = {Chemical Physics Letters},
	author = {Keith, T.A. and Bader, R.F.W.},
	month = jun,
	year = {1992},
	pages = {1--8},
}

@article{PhysRevLett.114.066404,
  title = {Gauge-Invariant Calculation of Static and Dynamical Magnetic Properties from the Current Density},
  author = {Raimbault, Nathaniel and de Boeij, Paul L. and Romaniello, Pina and Berger, J. A.},
  journal = {Phys. Rev. Lett.},
  volume = {114},
  issue = {6},
  pages = {066404},
  numpages = {5},
  year = {2015},
  month = {Feb},
  publisher = {American Physical Society},
  doi = {10.1103/PhysRevLett.114.066404},
  url = {https://link.aps.org/doi/10.1103/PhysRevLett.114.066404}
}

@Article{cheeseman_comparison_1996,
	author   = {Cheeseman, James R. and Trucks, Gary W. and Keith, Todd A. and Frisch, Michael J.},
	journal  = {J. Chem. Phys.},
	title    = {A comparison of models for calculating nuclear magnetic resonance shielding tensors},
	year     = {1996},
	issn     = {0021-9606, 1089-7690},
	month    = apr,
	number   = {14},
	pages    = {5497--5509},
	volume   = {104},
	doi      = {10.1063/1.471789},
	url      = {http://aip.scitation.org/doi/10.1063/1.471789},
	urldate  = {2021-10-02},
}

@article{keith_topological_1993,
	title = {Topological Analysis of Magnetically Induced Molecular Current Distributions},
	volume = {99},
	issn = {0021-9606, 1089-7690},
	url = {http://aip.scitation.org/doi/10.1063/1.466165},
	doi = {10.1063/1.466165},
	number = {5},
	urldate = {2018-08-07},
	journal = {J. Chem. Phys.},
	author = {Keith, Todd A. and Bader, Richard F. W.},
	month = sep,
	year = {1993},
	pages = {3669--3682}
}

@Article{becke_multicenter_1988,
	author   = {Becke, A. D.},
	journal  = {J. Chem. Phys.},
	title    = {A multicenter numerical integration scheme for polyatomic molecules},
	year     = {1988},
	issn     = {0021-9606, 1089-7690},
	month    = feb,
	number   = {4},
	pages    = {2547--2553},
	volume   = {88},
	doi      = {10.1063/1.454033},
	url      = {http://aip.scitation.org/doi/10.1063/1.454033},
	urldate  = {2020-01-14},
}

@Article{Keith_calculation_1993,
	author   = {Todd A. Keith and Richard F.W. Bader},
	journal  = {Chem. Phys. Lett.},
	title    = {Calculation of magnetic response properties using a continuous set of gauge transformations},
	year     = {1993},
	issn     = {0009-2614},
	number   = {1},
	pages    = {223-231},
	volume   = {210},
	doi      = {https://doi.org/10.1016/0009-2614(93)89127-4},
	url      = {https://www.sciencedirect.com/science/article/pii/0009261493891274},
}

@Article{monaco_program_2021,
  title = {Program Package for the Calculation of Origin-Independent Electron Current Density and Derived Magnetic Properties in Molecular Systems},
volume = {61},
ISSN = {1549-960X},
url = {http://dx.doi.org/10.1021/acs.jcim.0c01136},
DOI = {10.1021/acs.jcim.0c01136},
number = {1},
journal = {Journal of Chemical Information and Modeling},
publisher = {American Chemical Society (ACS)},
author = {Monaco,  Guglielmo and Summa,  Francesco F. and Zanasi,  Riccardo},
year = {2020},
month = Dec,
pages = {270–283}
}

@Article{summa_assessment_2021,
  title = {Assessment of the performance of DFT functionals in the fulfillment of off-diagonal hypervirial relationships},
volume = {23},
ISSN = {1463-9084},
url = {http://dx.doi.org/10.1039/D1CP01298C},
DOI = {10.1039/d1cp01298c},
number = {28},
journal = {Physical Chemistry Chemical Physics},
publisher = {Royal Society of Chemistry (RSC)},
author = {Summa,  Francesco F. and Monaco,  Guglielmo and Lazzeretti,  Paolo and Zanasi,  Riccardo},
year = {2021},
pages = {15268–15274}
}

@article{pennanen_density_2005,
 title = {Density-functional calculations of relativistic spin-orbit effects on nuclear magnetic shielding in paramagnetic molecules},
volume = {123},
ISSN = {1089-7690},
url = {http://dx.doi.org/10.1063/1.2079947},
DOI = {10.1063/1.2079947},
number = {17},
journal = {The Journal of Chemical Physics},
publisher = {AIP Publishing},
author = {Pennanen,  Teemu O. and Vaara,  Juha},
year = {2005},
month = Oct 
}

@article{bouten_relativistic_2000,
  title = {Relativistic Effects for NMR Shielding Constants in Transition Metal Oxides Using the Zeroth-Order Regular Approximation},
volume = {104},
ISSN = {1520-5215},
url = {http://dx.doi.org/10.1021/jp994480w},
DOI = {10.1021/jp994480w},
number = {23},
journal = {The Journal of Physical Chemistry A},
publisher = {American Chemical Society (ACS)},
author = {Bouten,  R. and Baerends,  Evert-Jan and van Lenthe,  Erik and Visscher,  Lucas and Schreckenbach,  Georg and Ziegler,  Tom},
year = {2000},
month = May,
pages = {5600–5611}
}

@article{van_lenthe_zero-order_1996,
 title = {The zero-order regular approximation for relativistic effects: The effect of spin–orbit coupling in closed shell molecules},
volume = {105},
ISSN = {1089-7690},
url = {http://dx.doi.org/10.1063/1.472460},
DOI = {10.1063/1.472460},
number = {15},
journal = {The Journal of Chemical Physics},
publisher = {AIP Publishing},
author = {van Lenthe,  Erik  and Snijders,  Jaap G. and Baerends,  Evert-Jan},
year = {1996},
month = Oct,
pages = {6505–6516}
}

@article{CDT,
  title = {Frequency-dependent current density tensors as density functions of dynamic polarizabilities},
volume = {150},
ISSN = {1089-7690},
url = {http://dx.doi.org/10.1063/1.5097578},
DOI = {10.1063/1.5097578},
number = {18},
journal = {The Journal of Chemical Physics},
publisher = {AIP Publishing},
author = {Lazzeretti,  Paolo},
year = {2019},
month = May 
}

@misc{x2c:2007,
note={{The generic acronym X2C (pronounced as ‘‘ecstacy’’) for exact two-component Hamiltonians resulted from intensive discussions among H. J. Aa.
Jensen, W. Kutzelnigg, W. Liu, T. Saue and L. Visscher during the Twelfth International Conference on the Applications of Density Functional Theory
(DFT-2007), Amsterdam, 26–30 August 2007. Note that the ‘‘exact’’ here means only that all the solutions of the Dirac-based Hamiltonian can be
reproduced up to machine accuracy. It is particularly meaningful when compared with finite order quasirelativistic theories.}}}

@article{Novotny2024,
  title = {{Paramagnetic Effects in NMR Spectroscopy of Transition-Metal Complexes: Principles and Chemical Concepts}},
  volume = {57},
  ISSN = {1520-4898},
  url = {http://dx.doi.org/10.1021/acs.accounts.3c00786},
  DOI = {10.1021/acs.accounts.3c00786},
  number = {10},
  journal = {Accounts of Chemical Research},
  publisher = {American Chemical Society (ACS)},
  author = {Novotny,  Jan and Komorovsky,  Stanislav and Marek,  Radek},
  year = {2024},
  month = Apr,
  pages = {1467–1477}
}

@article{Yuan2024,
  title = {{Formulation and Implementation of Frequency-Dependent Linear Response Properties with Relativistic Coupled Cluster Theory for GPU-Accelerated Computer Architectures}},
  volume = {20},
  ISSN = {1549-9626},
  url = {http://dx.doi.org/10.1021/acs.jctc.3c00812},
  DOI = {10.1021/acs.jctc.3c00812},
  number = {2},
  journal = {Journal of Chemical Theory and Computation},
  publisher = {American Chemical Society (ACS)},
  author = {Yuan,  Xiang and Halbert,  Loïc and Pototschnig,  Johann Valentin and Papadopoulos,  Anastasios and Coriani,  Sonia and Visscher,  Lucas and Pereira Gomes,  André Severo},
  year = {2024},
  month = Jan,
  pages = {677–694}
}
\end{document}